\documentclass[aps,prd,superscriptaddress,nofootinbib,amsmath,amsfonts,preprintnumbers,notitlepage,longbibliography,10pt,english]{revtex4-1}
\usepackage{amsmath}
\usepackage{amssymb}
\usepackage{babel}

\makeatletter

\usepackage{array,multirow,graphicx}
\usepackage{dcolumn}
\usepackage{newlfont}
\usepackage{bm}
\usepackage[colorlinks,citecolor=blue,urlcolor=blue,linkcolor=blue]{hyperref}
\usepackage[figtopcap]{subfigure}
\usepackage{color}

\newcommand{\arctanh}{\mathrm{arctanh}}

\begin{document}

\date{\today}

\title{Thermodynamical correspondence of $f(R)$ gravity in Jordan and Einstein frames}

\author{G.G.L. Nashed}
\email{nashed@bue.edu.eg}
\affiliation {Centre for Theoretical Physics, The British University, P.O. Box
43, El Sherouk City, Cairo 11837, Egypt}

\author{W. El Hanafy}
\email{waleed.elhanafy@bue.edu.eg}
\affiliation {Centre for Theoretical Physics, The British University, P.O. Box
43, El Sherouk City, Cairo 11837, Egypt}

\author{S.D. Odintsov}
\email{odintsov@ieec.uab.es}
\affiliation{Institut de Ci\`encies de l'Espai (ICE-CSIC/IEEC),\\
Campus UAB, c. Can Magrans s/n, 08193, Barcelona, Spain}
\affiliation{Instituci\'o Catalana de Recerca i Estudis Avan\c{c}ats (ICREA),
Barcelona, Spain}

\author{V.K. Oikonomou}
\email{v.k.oikonomou1979@gmail.com}
\affiliation{Department of Physics, Aristotle University of Thessaloniki, Thessaloniki 54124, Greece}
\affiliation{Laboratory for Theoretical Cosmology, Tomsk State University of Control Systems and Radioelectronics (TUSUR), 634050 Tomsk, Russia}

\begin{abstract}
We study the thermodynamical aspects of $f(R)$ gravity in the
Jordan and the Einstein frame, and we investigate the
corresponding equivalence of the thermodynamical quantities in the
two frames. We examine static spherically symmetric black hole
solutions with constant Ricci scalar curvature $R$, and as we
demonstrate, the thermodynamical quantities in the two frames are
equivalent. However, for the case of black holes with non-constant
scalar curvature $R$, the thermodynamical equivalence of the two
frames is no longer valid. In addition, we examine cosmological
solutions with non-trivial curvatures and as we demonstrate the
thermodynamical quantities in both frames are not equivalent. In
conclusion, although $f(R)$ gravity and its corresponding
scalar-tensor theory are mathematically equivalent, at least for
conformal invariant quantities, the two frames are not
thermodynamically equivalent at a quantitative level, in terms of
several physical quantities.

%\keywords{.}
%\pacs{}
\end{abstract}

\maketitle

\section{Introduction }\label{S1}

Gravity is one of the fundamental interactions in nature, which can be mathematically represented by the curvature of the spacetime as required by the general theory of relativity (GR). In this sense, the theory explains gravitation not as a long range force, but as a manifestation of the spacetime geometry itself. Indeed, this interpretation has been successfully verified on the solar system scale, the direct detection of gravitational waves by LIGO/VIRGO \cite{Abbott:2016blz,TheLIGOScientific:2016pea}, and even on the exotic level of the black hole horizon, it has been shown that the observed image of the shadow of a supermassive black hole at the center of the galaxy M87 is as predicted by general relativity, in particular Kerr black hole \cite{Akiyama:2019cqa,Akiyama:2019bqs}. However, on the cosmological scales, the confirmed observations of the late accelerated expansion and the galaxy formation lead to introduce some exotic matter components, dark matter and dark energy, if one considers the general relativity as the true theory of gravity. On the other hand, a reasonable alternative of the dark components is to modify gravity. One of the most prominent theory is $f(R)$ gravity \cite{Nojiri:2003ft,Nojiri:2010wj,Nojiri:2017ncd,Capozziello:2019cav,Capozziello:2011et,delaCruzDombriz:2012xy,Capozziello:2003tk,
Carroll:2003wy,Faraoni:2017afs}, which can be done by replacing the Ricci scalar, $R$, in Einstein--Hilbert action by an $f(R)$ function.

This class of theories shows an interesting mathematical feature, that is the adopted $f(R)$ theory in Jordan frame can be always transformed to a corresponding scalar--tensor theory in Einstein frame via a conformal transformation. This fact gives an evidence on the  mathematical equivalence of Jordan and Einstein frames \cite{Clifton:2008jq}. However, this equivalence cannot be extended to the physical level as pointed out in several works, c.f. \cite{Magnano:1993bd,Capozziello:2005mj,Capozziello:2006dj,Briscese:2006xu} (see also \cite{Faraoni:1999hp}). In more detail, the physical inequivalence of the two frames has been addressed in \cite{Dabrowski:2008kx}, whereas a vacuum solution in one frame conformally transforms to the other frame in presence of matter, this leads to conclude that conformal transformation can create matter, and therefore the two frames are not physically equivalent. Moreover, it has been shown that the singularity type changes when when one moves from frame to another \cite{Bahamonde:2016wmz}, even more the conformal transformation of anisotropic singularities in Jordan frame produces solutions free from singularities in Einstein frame \cite{Chakraborty:2018ost}. Moreover, the study of the thermodynamics of $f(R)$ gravity with disformal transformation shows that the non-equilibrium description in the Jordan frame cannot be extended to the Einstein frame, which leads to the conclusion that the two frames are not physically equivalent \cite{Geng:2019wgd}. On the contrary, the cosmic background radiation has been verified to be conformally invariant in the two frames\cite{Antoniadis:1996dj}. Furthermore, it has been shown that the slow-roll parameters are invariant quantities \cite{Kuusk:2016rso}. From a theoretical point of view, the conformal transformation is completely reasonable however the physical correspondence between the two frames is still debatable \cite{1997CQGra..14.2963C,Capozziello:1996xg,Faraoni:1998qx,1988PhLB..214..515B,Karam:2019dlv,Karam:2018squ,
Karam:2017zno,Bahamonde:2016wmz,Kaiser:1995nv,Ruf:2017xon}. It is the aim of the present study to investigate the equivalence between the two frames of $f(R)$ gravity, namely in the Einstein and Jordan frame, using black hole solutions and cosmological models on the thermodynamic level. In principle, the equivalence of the two frames can be granted for conformal invariant quantities, so the rest of the quantities which are not conformal invariant, should be explicitly studied with regard to their equivalence in the two frames.

The outline of the paper is: In Section \ref{S2}, a brief account of $f(R)$--Maxwell theory in Jordan frame and its conformal scalar--tensor theory in Einstein frame is presented. In Section \ref{S3}, for a spherically symmetric black hole solution with a constant curvature, the correspondence of $f(R)$ gravity in Jordan and Einstein frames is critically discussed. In Section \ref{S4}, for a spherically symmetric black hole solution with a non-constant curvature, the correspondence of $f(R)$ gravity in Jordan and Einstein frames is examined. We extend our study to include cosmological models with homogeneous and isotropic symmetry of flat FLRW solution. In Section \ref{S5}, the correspondence of $f(R)$ gravity, in power-law cosmology, in Jordan and Einstein frames is investigated. In Section \ref{S6}, the correspondence in a cosmological solution other than the power-law cosmology is examined. In Section \ref{S7}, the conclusions follow.

\section{Field Equations of $f(R)$--Maxwell Theory in the Jordan and the Einstein frames}\label{S2}

In this section, we recall some essential information about representation of $f(R)$ gravity in Jordan and Einstein frames. Thus, in absence of ordinary matter, the action of the Jordan frame $f(R)$ gravity coupled to Maxwell field is, \cite{1970MNRAS.150....1B,Nojiri:2010wj,Nojiri:2017ncd,Carroll:2003wy}
\begin{eqnarray} \label{J-action1}
\mathcal{S}_J=\mathcal{ S}_g+\mathcal{ S}_{e.m.}\, .
\end{eqnarray}
Here, the $4$-dimensional gravitational action ${\mathop{\mathcal{ S}}}_g$ of an $f(R)$ theory in this frame is,
\begin{eqnarray} \label{fR-action}
{\mathop{\mathcal{ S}}}_g=\frac{1}{2\kappa} \int d^4x \sqrt{-g}\,
\left[f(R)-\Lambda\right]\, ,
\end{eqnarray}
where $\Lambda$ is the cosmological constant, $R$ is the Ricci scalar, $\kappa=8\pi G=8\pi/m_p^2$ (with $G$ is Newton's gravitational constant and $m_p =1.22\times 10^{19}$ GeV is Planck's mass), $g$ is the determinant of the metric and $f(R)$ is an analytic differentiable function of the Ricci scalar $R$. In addition, the electrodynamics action ${\mathop{\mathcal{S}}}_{e.m.}$ is,
\begin{eqnarray}\label{a3}
{\mathop{\mathcal{S}}}_{e.m.}=-\frac{1}{2}F^{2}=-\frac{1}{2}F_{\mu
\nu}F^{\mu\nu}\, ,
\end{eqnarray}
where $F_{\mu \nu} =2A_{[\mu, \nu]}$ is the anti-symmetric electromagnetic tensor\footnote{The comma denotes the ordinary differentiation, the square bracket represents anti-symmetrization, i.e. $A_{[\mu, \nu]}=\frac{1}{2}(A_{\mu, \nu}-A_{\nu ,\mu})$ and the symmetrization is represented as $A_{(\mu, \nu)}=\frac{1}{2}(A_{\mu, \nu}+A_{\nu ,\mu})$ \cite{Hassaine:2007py}.}, and $A_\mu$ denotes the 1-form gauge potential.

The variation of the action (\ref{J-action1}) with respect to the metric tensor $g_{\mu \nu}$ and the electromagnetic field strength $F$ gives, respectively, the following set of field equations \cite{Cognola:2005de},
\begin{eqnarray} \label{J-field-eqn1}
I_{\mu \nu}=R_{\mu \nu} f_R-\frac{1}{2}g_{\mu \nu}f(R)-2g_{\mu \nu}\Lambda +g_{\mu \nu} \Box f_R-\nabla_\mu \nabla_\nu f_R-\kappa T_{\mu \nu}\equiv0\, ,
\end{eqnarray}
\begin{equation}\label{J-field-eqn2}
\partial_\nu \left( \sqrt{-g}\, F^{\mu \nu}\right)=0\, ,
\end{equation}
where $f_R=\frac{\partial f}{\partial R}$ and $T_{\mu \nu}$ defines the traceless energy--momentum tensor of the electrodynamics field,
\begin{equation}
T_{\mu \nu}=\frac{1}{\kappa}\left(2g_{\rho\sigma} F_\nu{^\rho}
F_\mu{^\sigma}-\frac{1}{2} g_{\mu \nu} F^{2}\right)\, .
\end{equation} By Taking the trace of equation
(\ref{J-field-eqn1}), we have,
\begin{eqnarray} \label{J-field-eqn3}
Rf_R-2f(R)-8\Lambda+3\Box f_R=0\, .
\end{eqnarray}
It has been shown that the $f(R)$ gravity can be written in the form of the Brans-Dicke theory by introducing an auxiliary field $\chi$ through a non-minimal coupling term as in the following action,
\begin{equation}\label{J-action2}
\mathcal{S}_J= \int d^4x \sqrt{-g}\, \left[\frac{1}{2\kappa}
f_\chi(\chi)(R-\Lambda)-\underbrace{\left(\frac{\chi
f_\chi(\chi)-f(\chi)}{2\kappa}\right)}_{V_J(\chi)}\right]+\mathcal{S}_{e.m.}
\, .
\end{equation}
The variation of the action with respect to $\chi$ yields $f_{\chi\chi}(R-\chi)=0$. For $f_{\chi\chi}\neq 0$, i.e. $\chi=R$, the above action turns back to the action (\ref{J-action1}). In this sense, the field equations produced by the action (\ref{J-action2}) are identical to those previously obtained from action (\ref{J-action1}), namely equations (\ref{J-field-eqn1}) and (\ref{J-field-eqn2}).

By choosing $\sigma=f_{\chi}(\chi)$, the action (\ref{J-action2}) reads as Brans--Dicke like theory with a non-minimal coupling term $\sigma R$ and a scalaron potential $V(\sigma)$. However, one can always eliminate the non-minimal coupling term in the Jordan frame by moving to the Einstein frame via the conformal transformation,
\begin{eqnarray} \label{conf-trans}
g_{\mu \nu} \to {\bar g}_{\mu \nu}(x)=\Omega^2(x) g_{\mu \nu}(x)\, ,
\end{eqnarray}
with the spacetime conformal factor $\Omega$ chosen as $\Omega^2(x) =f_R$, which requires $f_R > 0$ \cite{Bahamonde:2017kbs,Bahamonde:2016wmz}. We note that under the transformation (\ref{conf-trans}), the Ricci scalar transforms as $R\to \bar{R}$. However, by introducing the canonical scalar field,
\begin{equation}\label{scalar-field}
\phi=\sqrt{\frac{6}{\kappa}}\, \ln \Omega =\sqrt{\frac{3}{2\kappa}}\, \ln f_R \, ,
\end{equation}
and by applying the conformal transformation (\ref{conf-trans}), the action (\ref{J-action2}) transforms to the corresponding scalar-tensor theory in the Einstein frame \cite{Bahamonde:2017kbs},
\begin{eqnarray} \label{E-action1}
{\mathop{\mathcal{ S}_E}}= \int d^4x \sqrt{-{\bar g}}
\left[\frac{1}{2\kappa}({\bar R}-\Lambda)-\frac{1}{2}{\bar g}^{\mu
\nu}\partial_\mu \phi \, \partial_\nu
\phi-V_E(\phi)\right]+\mathcal{S}_{e.m.}\, ,
\end{eqnarray}
where the potential of the canonical scalar field $\phi$ reads,
\begin{equation}\label{pot-fR}
V_E(\phi)=\displaystyle\frac{Rf_R-f}{2\kappa f_R^2}=\frac{V_J(\chi)}{f_R^2}\, .
\end{equation}
We note that the potential $V_E(\phi)$ can be expressed in terms of $\phi$ using the inverse relation $f_R=e^{ \sqrt{2\kappa/3}\,\phi}$. In addition, the energy--momentum tensor transforms as
$$T_{\mu\nu} \to \bar{T}_{\mu\nu}=\Omega(x)^{-2} T_{\mu\nu}.$$

We close this section mentioning that the physical interpretation of the correspondence of Jordan and Einstein frames is not an easy task. The equivalence of the two frames is supported by several studies, for example the equivalence of the inflationary parameters in both frames \cite{Brooker:2016oqa}, also the thermodynamical equivalence has been discussed in the case of BD theory \cite{Bhattacharya:2017pqc}. On the contrary, some work shows that the thermodynamic on cosmic scale is inequivalent in the two frames in the disformal $f(R)$ gravity \cite{Geng:2019wgd}. In addition, the difference of the two frames has been shown in other cases, c.f. \cite{Capozziello:2006dj,Capozziello:2010sc}. Moreover, for the case that the finite time singularities could transform from one type in one frame to another type in the other frame \cite{Bahamonde:2016wmz}. Furthermore, it has been shown that $f(R)$ models free from anisotropic singularities could be constructed in Einstein frame, but it is not necessarily true that these models are non-singular in the Jordan frame \cite{Chakraborty:2018ost}. In the same line of research, we focus in this study on some thermodynamical quantities which may give a new perspective in the Jordan-Einstein frame (non-)equivalence.

%%%%%%%%%%%%%%%%%%%%%%%%%%%%%%%%%%%%%%%%%%%%%%%%%%%%%%%%%%%%%%%%%%%%%%%%%%%%%%%%%%%%%%%%here

\section{Black holes with Constant Curvature in $f(R)$ Gravity}\label{S3}

In this section, we discuss the thermodynamics of spherically symmetric black hole solutions of $f(R)$ gravity in both Jordan and Einstein frames, focusing on those black holes which have constant scalar curvature. The aim is to investigate the validity of the correspondence of these solutions on the thermodynamics level.

\subsection{Thermodynamics of Black Holes in the Jordan frame}\label{S3.1}

\begin{figure}
\centering
\subfigure[~Possible two horizons]{\label{fig:1a}\includegraphics[scale=0.3]{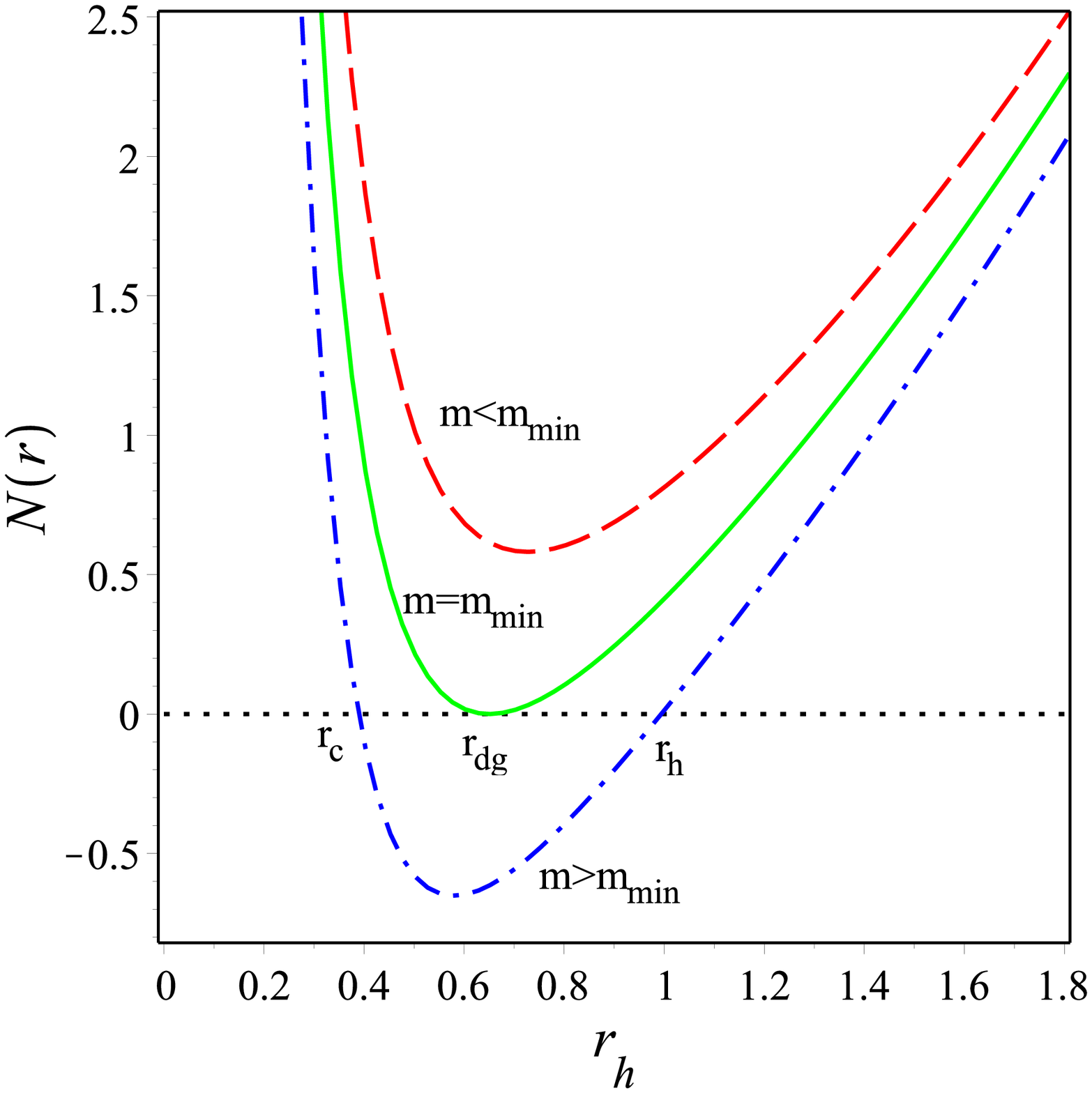}}
\subfigure[~The horizon mass-radius]{\label{fig:1b}\includegraphics[scale=0.3]{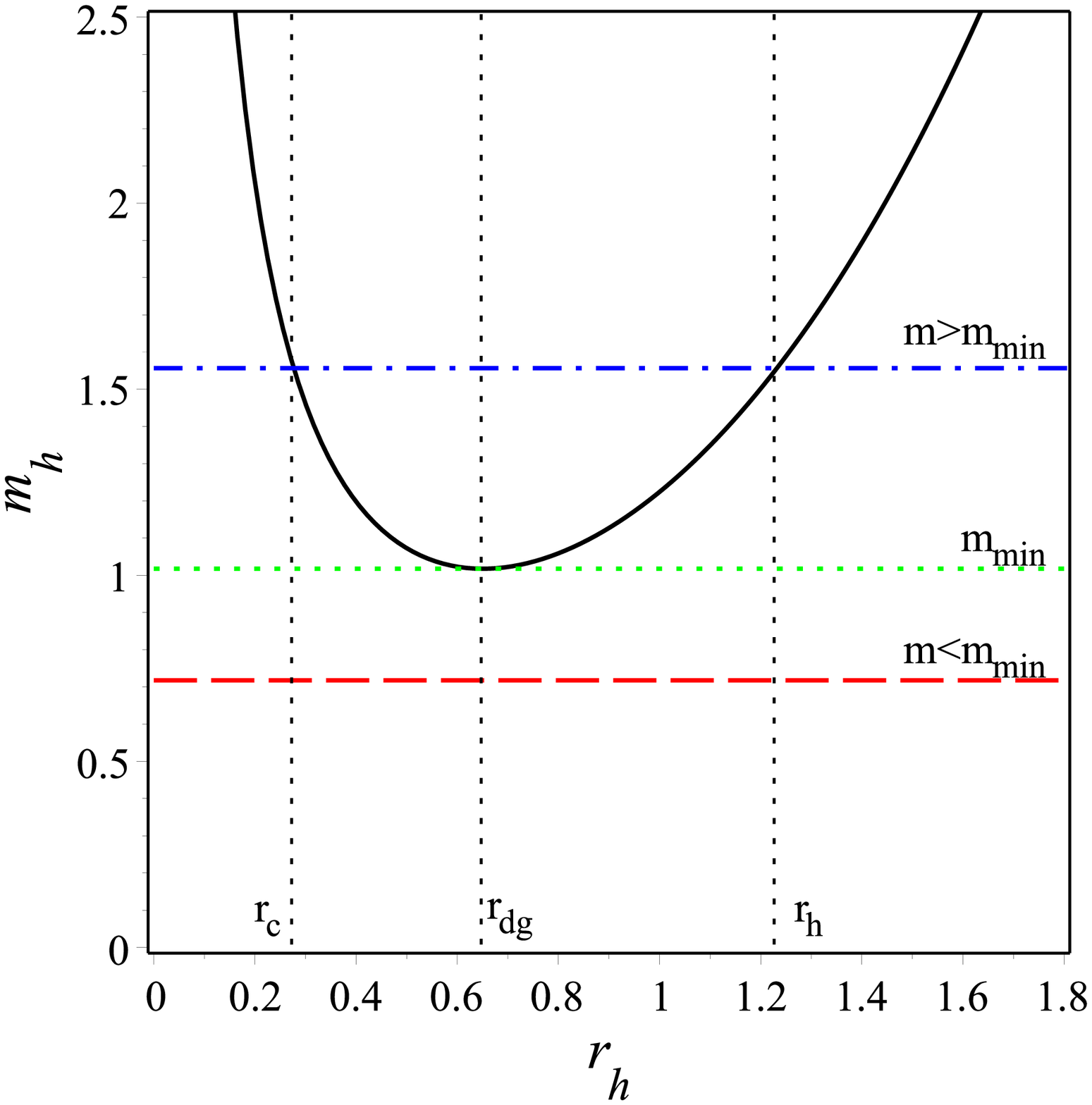}}
\subfigure[~The horizon Bekenstein-Hawking entropy]{\label{fig:1c}\includegraphics[scale=0.3]{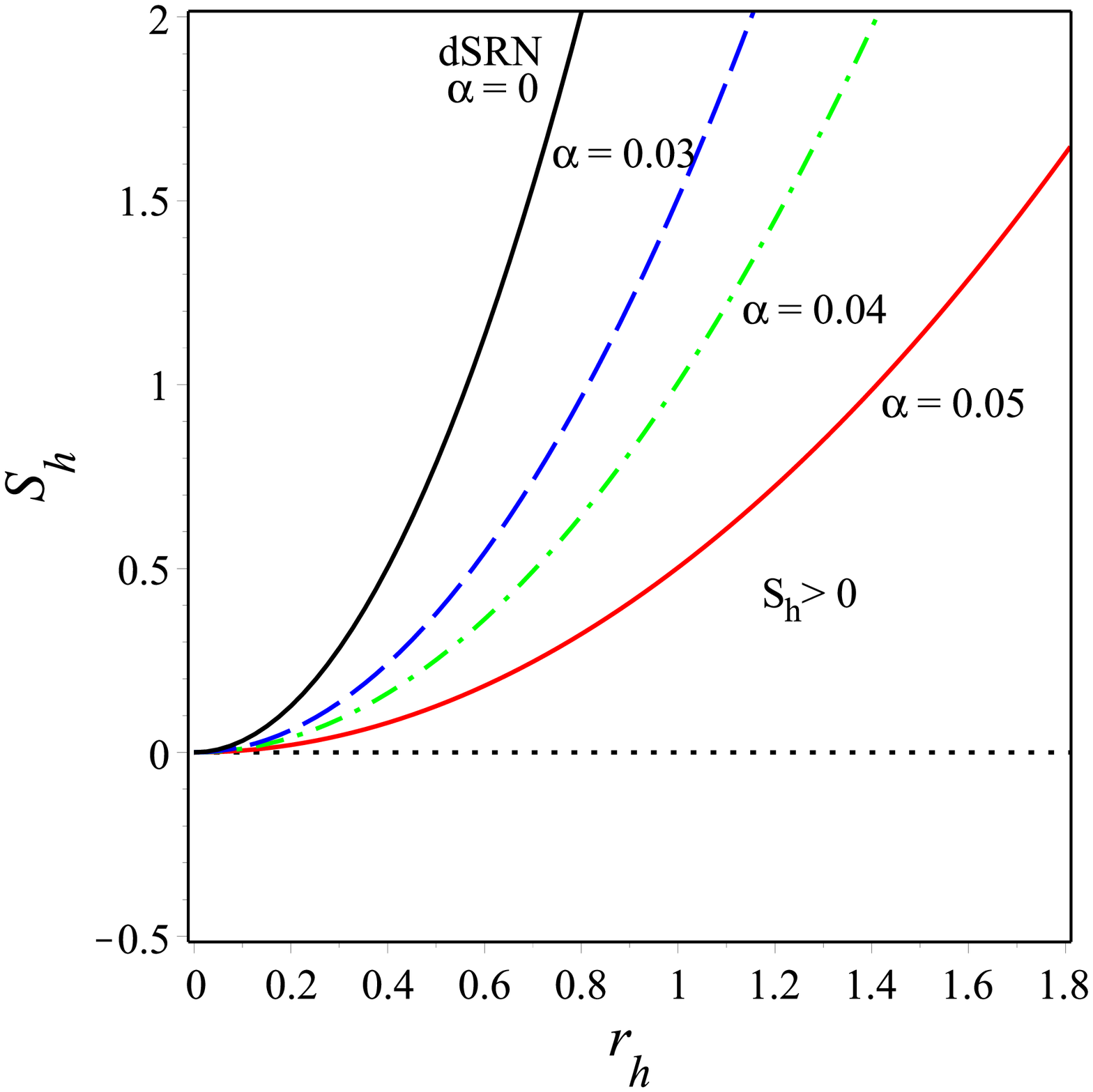}}\\
\subfigure[~The horizon Hawking Temperature]{\label{fig:1d}\includegraphics[scale=0.3]{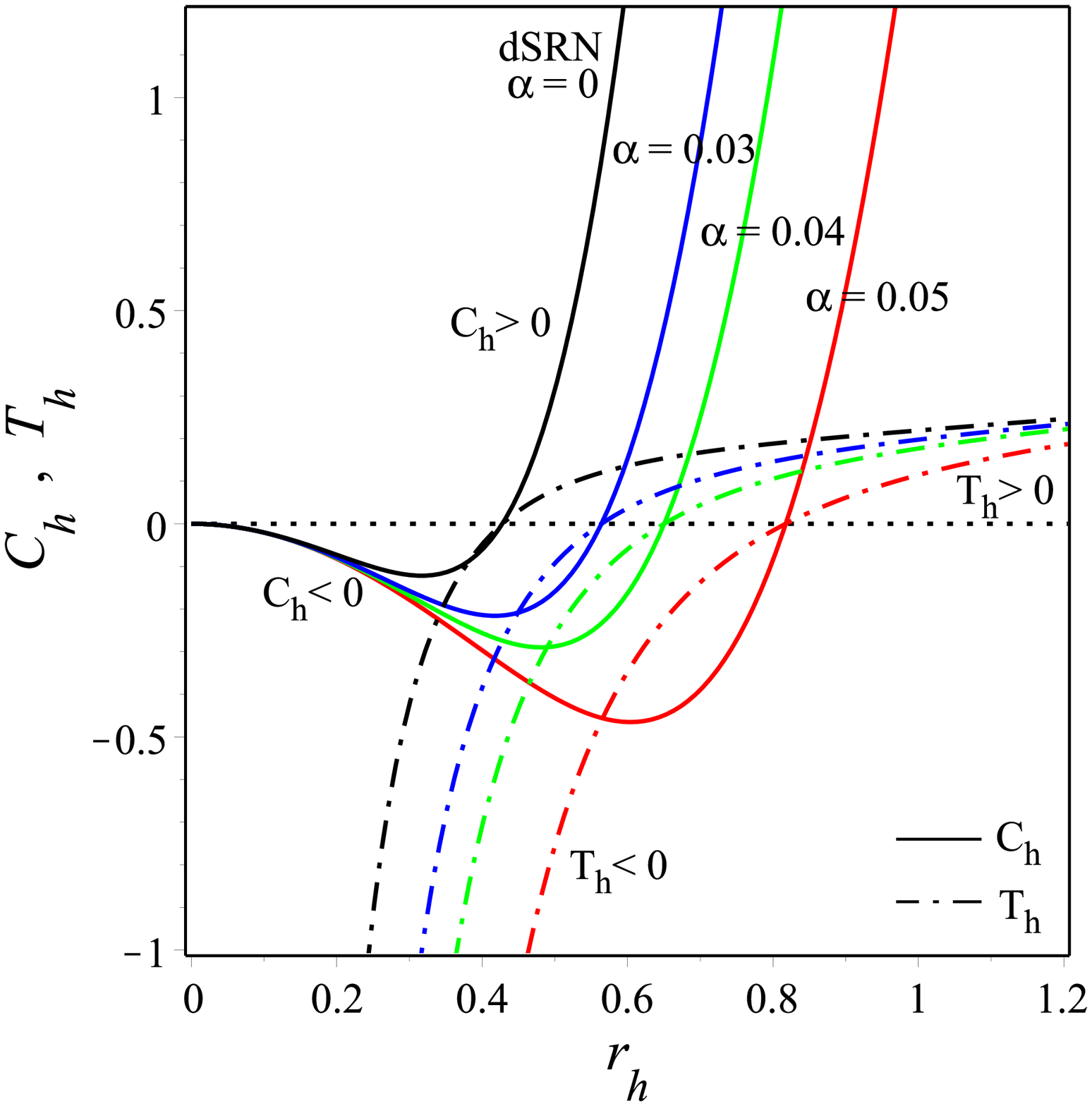}}
\subfigure[~The horizon quasi-local energy]{\label{fig:1e}\includegraphics[scale=0.3]{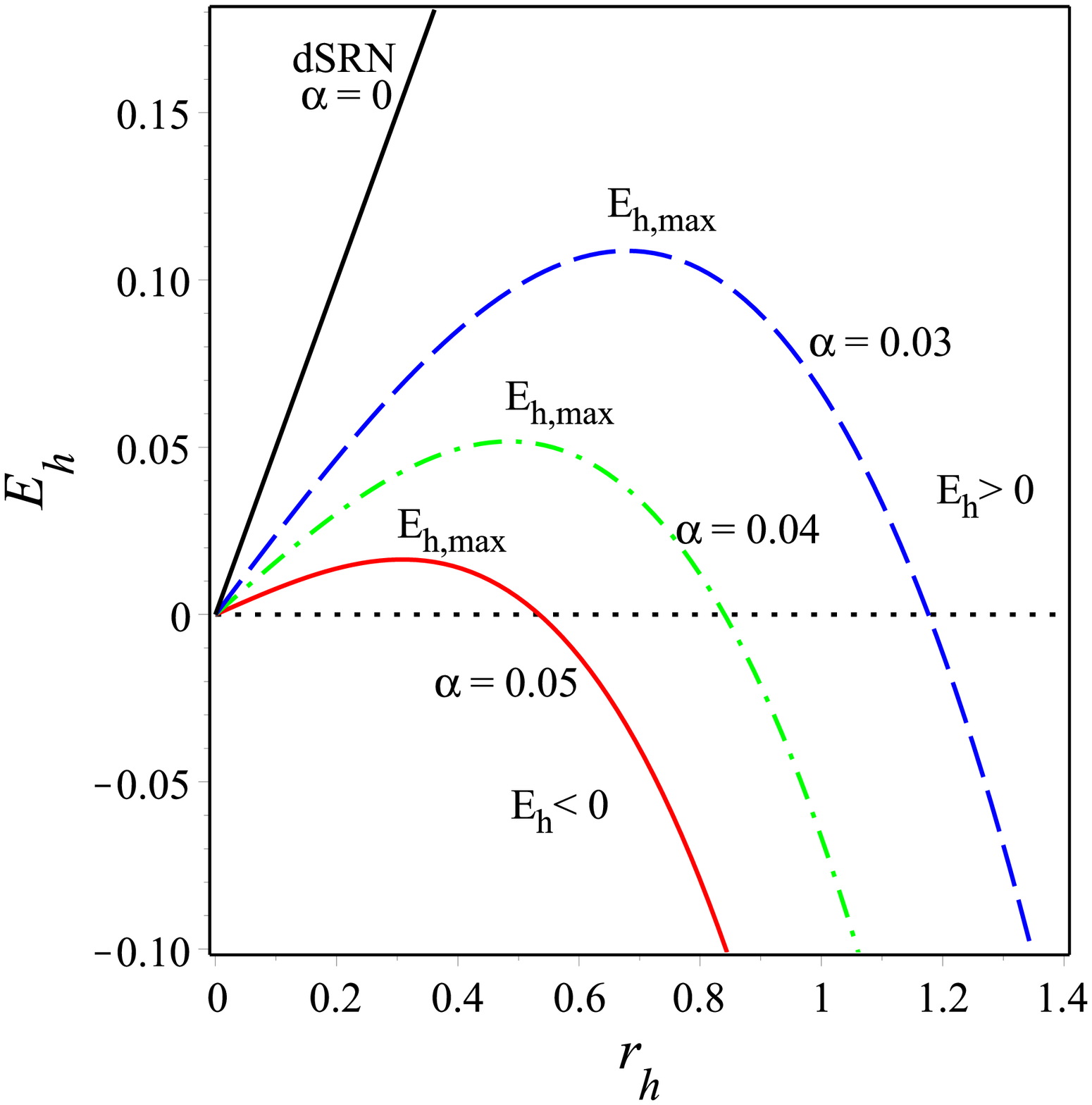}}
\subfigure[~The horizon Gibbs free energy]{\label{fig:1f}\includegraphics[scale=0.3]{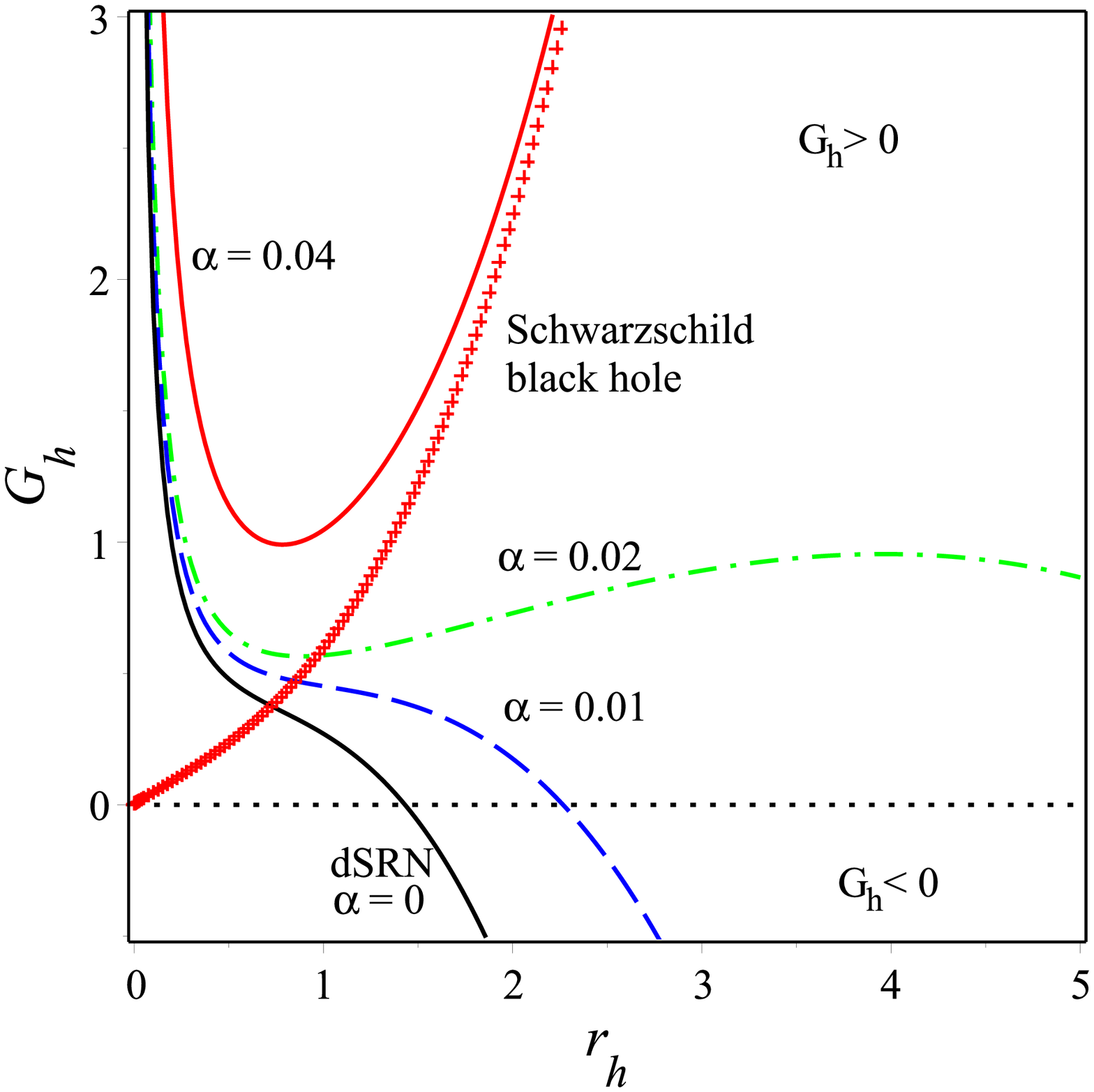}}
\caption{Schematic plots of thermodynamical quantities of the black hole solution (\ref{sol1a}) in Jordan frame: \subref{fig:1a} Typical behavior of the metric function $N(r)$ given by (\ref{sol1a}); \subref{fig:1b} the horizon mass--radius relation (\ref{hor-mass-rad1a}) with minimum mass given by (\ref{m_min1a}), it shows that, for $\Lambda>0$ case, the solution admits at most two horizons ($r_c$, $r_h$) when $m_h>m_{min}$, one degenerate horizon ($r_c=r_h=r_{dg}$) when $m_h=m_{min}$ and otherwise it provides a naked singularity; \subref{fig:1c} typical behavior of the horizon entropy, (\ref{BH-entropy1a}), which shows that $S_h$ increases quadratically as $r_h$ increases; \subref{fig:1d} typical behavior of the horizon temperature, (\ref{H-temp1a}), and the heat capacity, (\ref{heat-cap1a}), which shows that both vanish at $r_{dg}$, \eqref{deg_horz1a}, the black hole is unstable (i.e $C_h<0$) when $r_h<r_{dg}$ and it is stable (i.e $C_h>0$) when $r_h>r_{dg}$; \subref{fig:1e} typical behavior of the horizon quasi-local energy, (\ref{loc-eng1a}), which shows that $E_h>0$ is max when $r_h=\frac{\Omega\sqrt{2\alpha}}{8\alpha\Lambda}$ and it vanishes at $r_h=0$ and $r_h=\frac{\Omega\sqrt{6\alpha}}{8\alpha\Lambda}$, otherwise $E_h$ becomes negative; \subref{fig:1f} typical behavior of the horizon Gibbs free energy, (\ref{G-eng1a}), which shows that $G_h$ could have only positive values for large $\alpha \lesssim 1/(16\Lambda)$, whereas $G_h$ interpolates between dSRN and de Sitter Schwarzschild black holes, otherwise it peaks up at some $r_h$ and then drops to negative values at large $r_h$ as in dSRN case. We take $q=0.5$, $\Lambda=1$ and different values of $\alpha$ satisfies $0<\alpha<1/(16\Lambda)$, whereas $\alpha=0$ reduces to dSRN black hole.}
\label{Fig:1}
\end{figure}

For a static spherically symmetric black hole, we assume that the line element of the spacetime metric is,
\begin{equation}\label{line_element_1}
    ds_J^2= -N(r) dt^2+\frac{dr^2}{N(r)}+r^2 \left(d\theta^2+\sin^2 \theta d\varphi^2\right),
\end{equation}
where $-\infty < t < \infty$, $r \geq 0$, $0 \leq \theta < 2\pi$ and $0\leq \varphi \leq \pi$. We insert the metric (\ref{line_element_1}) in the $f(R)$ gravity field equations (\ref{J-field-eqn1}), and we take the $f(R)$ gravity to be,
\begin{equation}\label{fR1}
f(R)=R+\alpha R^2\, .
\end{equation}
By solving the equations of motion, we get the solution \cite{Nashed:2018efg},
\begin{equation} \label{sol1a}
 N(r)=1-\frac{2m}{r}+\frac{2\Lambda r^2}{3}+\frac{q^2}{\Omega^2r^2}\, ,
\end{equation}
where $\Omega=\sqrt{1-16\alpha \Lambda}$. It is easy to show that this solution leads to a constant Ricci scalar, $R=-8\Lambda$, as required for our study in this section. We note that the valid range of the parameter $\alpha$ is $0<\alpha<1/(16\Lambda)$ in order to satisfy the stability and the ghost free constraints $f_R>0$ and $f_{RR}>0$. For Anti de Sitter Reissner Nordstr\"om (AdSRN) spacetime, i.e $\Lambda<0$, Eq.  (\ref{sol1a}) produces three horizons one of them is  cosmological while the others two are event horizons. For de Sitter Reissner Nordstr\"om (dSRN) spacetime, i.e $\Lambda>0$, the solution  produces only two event horizons. The later is shown in Fig. \ref{Fig:1}\subref{fig:1a}, the plot shows that the black hole could have at most two horizons at the roots of $N(r)=0$, those are the Cauchy (inner) horizon $r_c$ and the event (outter) horizon $r_h$. We note that, in the case $m>0$, $q>0$ and $\Lambda>0$, we find that these two roots are possible as long as the black hole mass $m> m_{min}$, where
\begin{equation}\label{m_min1a}
m_{min}=\frac{1}{6}\left(\frac{\Psi}{\Omega}+2\right)
\sqrt{\frac{\Psi-\Omega}{\Omega \Lambda}}\, ,
\end{equation}
where $\Psi=\sqrt{\Omega^2+8\Lambda q^2}$. This identifies the minimum horizon mass--radius relation of the black hole or the so-called Nariai black hole. Interestingly, in this case the two horizons coincide forming one horizon, $r_h=r_c=r_{dg}$, that is the degenerate horizon,
\begin{equation}\label{deg_horz1a}
r_{dg}=\frac{1}{2}\sqrt{\frac{\Psi-\Omega}{\Omega \Lambda}}.
\end{equation}
It is useful to calculate the total mass contained within the event horizon $r_h$. This can be done by setting $N(r_h) = 0$, and then we obtain the horizon mass-radius relation,
\begin{eqnarray} \label{hor-mass-rad1a}
 {m_h}=\frac{(3+2\Lambda r_h^2)\Omega^2 r_h^2+3q^2}{6\Omega^2 r_h}\, .
\end{eqnarray}
We plot the above relation in Fig. \ref{Fig:1}\subref{fig:1b}, whereas $m_h$ has always positive values, the black hole has two horizons if $m_h > m_{min}$ and only one horizon if $m_h=m_{min}$, otherwise, there is no horizon. These results are in agreement with Fig. \ref{Fig:1}\subref{fig:1a}. However, the horizon mass-radius relation shows that $m_h$ diverges as $r_h\to 0$ and $r_h\to \infty$. Notably, from Eq. (\ref{hor-mass-rad1a}) one can alternatively identify the degenerate horizon, $r_d$, by setting $\partial m_h/\partial r_h=0$. This gives $r_{dg}$ as previously obtained in Eq. (\ref{deg_horz1a}).
Next, we derive some thermodynamical quantities essential for the considerations to follow. The Bekenstein-Hawking entropy in the context of $f(R)$ gravity is \cite{Cognola:2005de,Brevik:2004sd},
\begin{equation}\label{BH-entropy}
S_h\equiv S(r_h)=\frac{1}{4} f_{R}\,A ,
\end{equation}
where $A$ is the area of the event horizon. In the GR limit $f_R=1$, we get $S_h \propto A$. For the solution (\ref{sol1a}), we write,
\begin{eqnarray} \label{BH-entropy1a}
{S_h}=\pi  \Omega^2 r_h^2= (1-16\alpha\Lambda)\pi r_h^2\, .
\end{eqnarray}
As it is clear, the above form reduces to the one corresponding to GR results when $\alpha=0$. Notably, since $R$ is constant in the quadratic $f(R)$ gravity, the entropy still fulfills $S_h \propto A$ just as in GR gravity however, we have $f_R \neq 1$. For $0<\alpha<\frac{1}{16\Lambda}$, we obtain the typical entropy behavior $S_h>0$ and increases with $r_h$. Accordingly, we plot Bekenstein-Hawking entropy in Fig. \ref{Fig:1}\subref{fig:1c}.

The Hawking temperature $T(r_h)$ of the quadratic $f(R)$ gravity at the event horizon \cite{Nojiri:2003ft} is,
\begin{equation}\label{Hawking-temp}
T_h\equiv T(r_h) = \frac{1}{4\pi}N'(r_h),
\end{equation}
where the prime denotes the derivative with respect to the the radial coordinates and the event horizon $r_h$ is the largest positive root of $N(r_h) = 0$ that satisfies $N'(r_h)\neq 0$. Substituting the solution (\ref{sol1a}) into Hawking temperature, we obtain,
\begin{eqnarray} \label{H-temp1a}
{T_h}=\frac{(1+2\Lambda r_h^2)\Omega^2 r_h^2-q^2}{4\pi \Omega^2 r_h^3}\, ,
\end{eqnarray}
Remarkably, if the event horizon, $r_h$, is located at the degenerate horizon, $r_{dg}$, namely (\ref{deg_horz1a}), the Hawking temperature vanishes. This can be shown in Fig. \ref{Fig:1}\subref{fig:1d}, where the temperature is negative if $r_h<r_{dg}$ and positive otherwise.

Notably, the stability of the black hole solution can be studied on the dynamical and the perturbative levels, c.f \cite{Nashed:2003ee,Myung:2011we,Myung:2013oca}, one of the essential quantities to investigate the thermodynamical stability of black holes is the heat capacity $C(r_h)$ at the event horizon. The event horizon heat capacity is given by \cite{Nouicer:2007pu,DK11,Chamblin:1999tk},
\begin{equation}\label{heat-capacity}
C_h\equiv C(r_h)=\frac{dE_h}{dT_h}= \frac{\partial m}{\partial r_h} \left(\frac{\partial T}{\partial r_h}\right)^{-1}\, .
\end{equation}
We note that the black hole is thermodynamically stable, if the heat capacity $C_h$ is positive, otherwise it is unstable. Substituting (\ref{hor-mass-rad1a}) and (\ref{H-temp1a}) into (\ref{heat-capacity}), we calculate the heat capacity,
\begin{equation}\label{heat-cap1a}
C_h=2\pi r_h^2 \frac{(2\Lambda r_h^2+1)\Omega^2 r_h^2-q^2}{\;\;(2 \Lambda r_h^2-1)\Omega^2 r_h^2+3q^2} \, .
\end{equation}
Since $C_h$ does not locally diverge as it is clear from the above equation, the black hole has no phase transition of second-order. We plot the heat capacity in Fig. \ref{Fig:1}\subref{fig:1d} which shows that $C_h<0 $ where $r_h<r_{dg}$ and the black hole is thermodynamically unstable. On the contrary, $C_h>0$ where $r>r_{dg}$ and the black hole is thermodynamically stable. Notably, the heat capacity vanishes on the degenerate horizon, $r_{dg}$, as well as the Hawking temperature.

Also, the quasi-local energy $E(r_h)$ of the $f(R)$ gravity (see \cite{Cognola:2011nj,Zheng:2018fyn}) at the event horizon is,
\begin{equation}\label{local-energy}
E_h\equiv E(r_h)=\frac{1}{4}\displaystyle{\int }\Bigg[2f_{R}(r_h)+r_h^2\Big\{f(R(r_h))-R(r_h)f_{R}(r_h)\Big\}\Bigg]dr_h,
\end{equation}
so for the quadratic $f(R)$ gravity (\ref{fR1}), we obtain
\begin{eqnarray} \label{loc-eng1a}
{E_h}=\frac{r_h}{2}\left[1-\frac{16}{3} \alpha \Lambda \left(3+2\Lambda r_h^2\right)\right]\, .
\end{eqnarray}
As it is clear, for $\alpha=0$ the Schwarzschild solution, $E_h=r_h/2$, is recovered. For $0<\alpha<1/(16\Lambda)$, the quasi-local energy vanishes at $r_h=0$ and also $r_h=\frac{\Omega\sqrt{6\alpha}}{8\alpha\Lambda}$, meanwhile it reaches a maximum value at $r_h=\frac{\Omega\sqrt{2\alpha}}{8\alpha\Lambda}$. This non trivial modification is due to the quadratic correction of the $f(R)$ gravity. We plot Eq. (\ref{loc-eng1a}) as in Fig. \ref{Fig:1}\subref{fig:1e}, the quasi-local energy is positive for $0 < r_h < \frac{\Omega\sqrt{6\alpha}}{8\alpha\Lambda}$, while it is negative for larger $r_h$.

Finally, the free energy in the grand canonical ensemble, which is also known as Gibbs free energy, is defined as, \cite{Zheng:2018fyn,Kim:2012cma}
\begin{equation} \label{Gibbs-energy}
G_h\equiv G(r_h)=m(r_h)-T(r_h)S(r_h).%+P(r_+)V(r_+),
\end{equation}
Using Eqs. (\ref{hor-mass-rad1a}), (\ref{BH-entropy1a}), (\ref{H-temp1a}) and (\ref{Gibbs-energy}), the Gibbs free energy reads,
\begin{eqnarray} \label{G-eng1a}
{G_h}=\frac{-3(1+2\Lambda r_h^2)r_h^2\Omega^4+\left[(3+2\Lambda r_h^2)r_h^2+3q^2\right]\Omega^2+6q^2}{12\Omega^2 r_h}\, .
\end{eqnarray}
In Fig.  \ref{Fig:1}\subref{fig:1f}, we plot Gibbs energy of the black hole (\ref{sol1a}) which shows that the Gibbs energy diverges at $r_h\to 0^+$ for non vanishing charge for all values of $0<\alpha<1/(16\Lambda)$. This is typical behavior of Gibbs energy of charged black holes which drop to negative values at large $r_h$. We note that for $\alpha=0$, the dSRN black hole is recovered. Interestingly, for large values of $\alpha \lesssim 1/(16\Lambda)$, Gibbs energy is always positive and interpolates between two black hole solutions, RdSN and Schwarzschild black holes. This clear in Fig.  \ref{Fig:1}\subref{fig:1f}, since $G_h$ diverges as $r_h\to 0^+$ as in dSRN case, then it drops to a minimum positive value and turns up to match Schwarzschild solution. So this case produces an intermediate case of these solutions.

As clear from the above calculations the thermodynamics of the black hole solution of the quadratic $f(R)$ gravity is non trivially modified. This is the case within Jordan frame, we next investigate the case within Einstein frame as well searching for possible correspondences on the thermodynamics level of these two frames, if any.

%%%%%%%%%%%%%%%%%%%%%%%%%%%%%%%%%%%%%%%%%%%%%%%%%%%%%%%%%%%%%%%%%%%%%%%%%%%%%%%%%%%%%%%%%%%%%%%%%h2

\subsection{Thermodynamics of Black Holes in the Einstein frame}\label{S3.2}

In this subsection, we reevaluate the same quantities that have been discussed in the previous subsection, but in the Einstein frame, investigating possible changes between the two frames. For that purpose, we apply the conformal transformation (\ref{conf-trans}) to the spacetime metric (\ref{line_element_1}), that is $d\bar{s}_E^2=\Omega^2 ds_J^2$, where the conformal factor of the $f(R)$ gravity (\ref{fR1}) is given by,
\begin{equation}\label{conf-trans1}
\Omega^2=f_R=1-16\alpha \Lambda\, .
\end{equation}
Thus, we write the Einstein frame metric as follows,
\begin{eqnarray}\label{line_element_1E}
\nonumber    d\bar{s}_E^2&=& \Omega^2\left[-N(r) dt^2+\frac{dr^2}{N(r)}+r^2 \left(d\theta^2+\sin^2 \theta d\varphi^2\right)\right]\, ,\\
    &=&-\bar{N}(\bar{r}) d\bar{t}^2+\frac{d\bar{r}^2}{\bar{N}(\bar{r})}+\bar{r}^2 \left(d\theta^2+\sin^2 \theta d\varphi^2\right)\, ,
\end{eqnarray}
where,
$$d\bar{t}=\Omega dt,\quad d\bar{r}=\Omega dr,\quad \bar{N}(\bar{r})=N(r(\bar{r})).$$
This gives the solution (\ref{sol1a}) in the Einstein frame, which is,
\begin{equation} \label{sol1b}
\bar{N}(\bar{r})=1-\frac{2 \bar{m}}{\bar{r}}+\frac{2 \bar{\Lambda} \bar{r}^2}{3}+\frac{\bar{q}^2}{\bar{r}^2}\, ,
\end{equation}
where $\bar{m}=\Omega\, m$, $\bar{q}=q$ and $\bar{\Lambda}=\Lambda/\Omega^2$. One can show that the above solution satisfies the field equations of the action (\ref{fR-action}). As it is clear, the solution (\ref{sol1b}) is just a re-scaled representation of the Jordan frame solution (\ref{sol1a}). This is a general feature of solutions with constant Ricci scalars so that no changes of the physical behavior of the thermodynamical quantities are expected. In other words, the correspondence of the $f(R)$ gravity, whose constant curvatures, in Jordan and Einstein frames holds true at the dynamical level as well as the thermodynamical one. This is due to the fact that any black hole solution with constant value of Ricci scalar is a GR solution, since the $f(R)$ field equations in this case are equivalent to GR with effective cosmological constant \cite{Cognola:2007zu,PhysRevD.74.086005}. Now the question is whether this true in the cases of the solutions with non-constant curvatures? This is the subject of the next section.

\section{Black Holes with non-constant Curvature in $f(R)$ gravity}\label{S4}

Similar to Section \ref{S3}, we discuss here the thermodynamics of spherically symmetric black hole solutions of an $f(R)$ gravity in both Jordan and Einstein frames, focusing on those black holes which have not constant scalar curvature. The aim of this section is to investigate the validity of the correspondence of these solutions, in both the Jordan and Einstein frame, at the thermodynamics level.

\subsection{Thermodynamics of Black Holes in the Jordan frame}\label{S4.1}

\begin{figure}
\centering
\subfigure[~Possible two horizons]{\label{fig:2a}\includegraphics[scale=0.3]{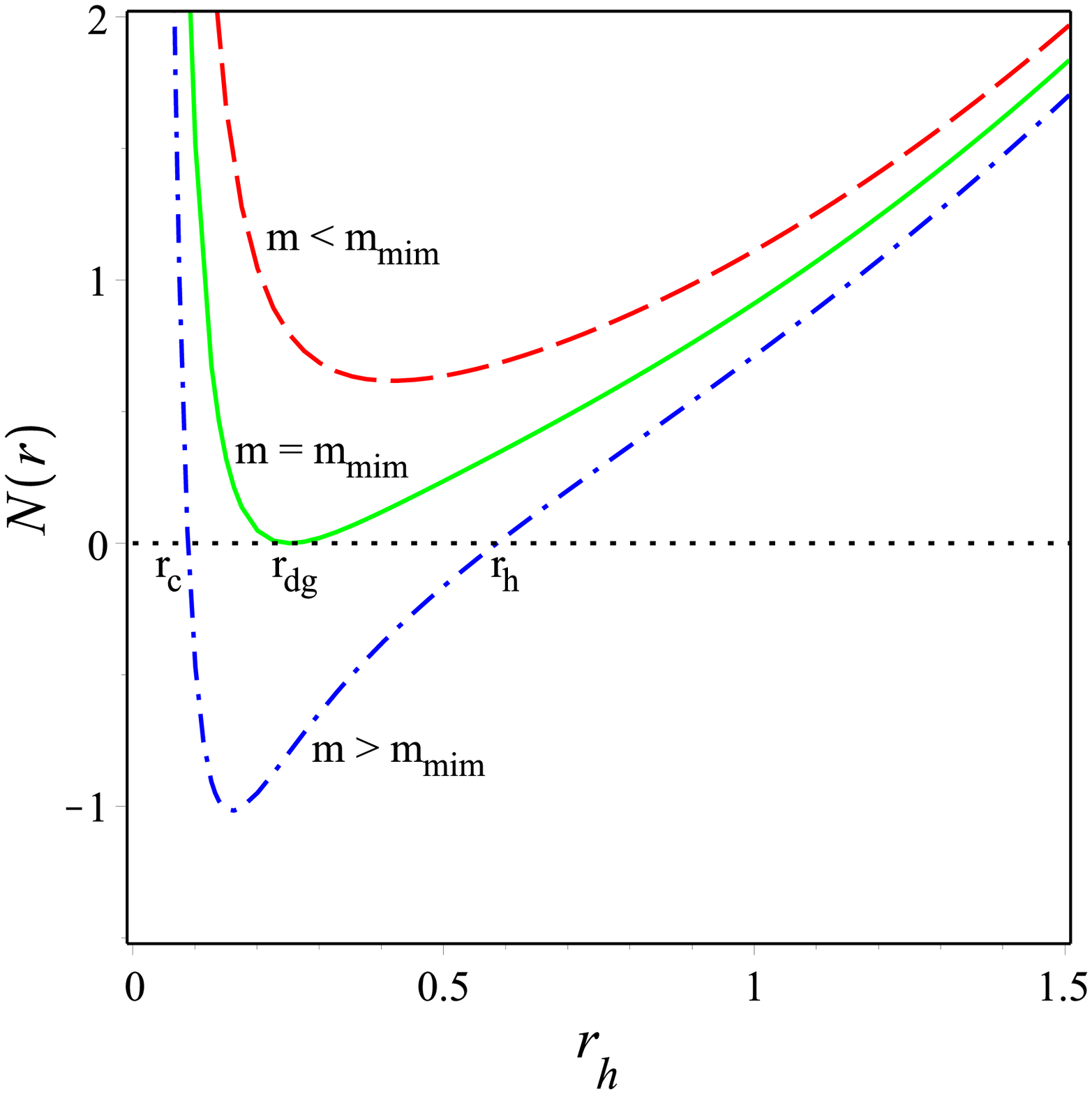}}
\subfigure[~The horizon mass-radius]{\label{fig:2b}\includegraphics[scale=0.3]{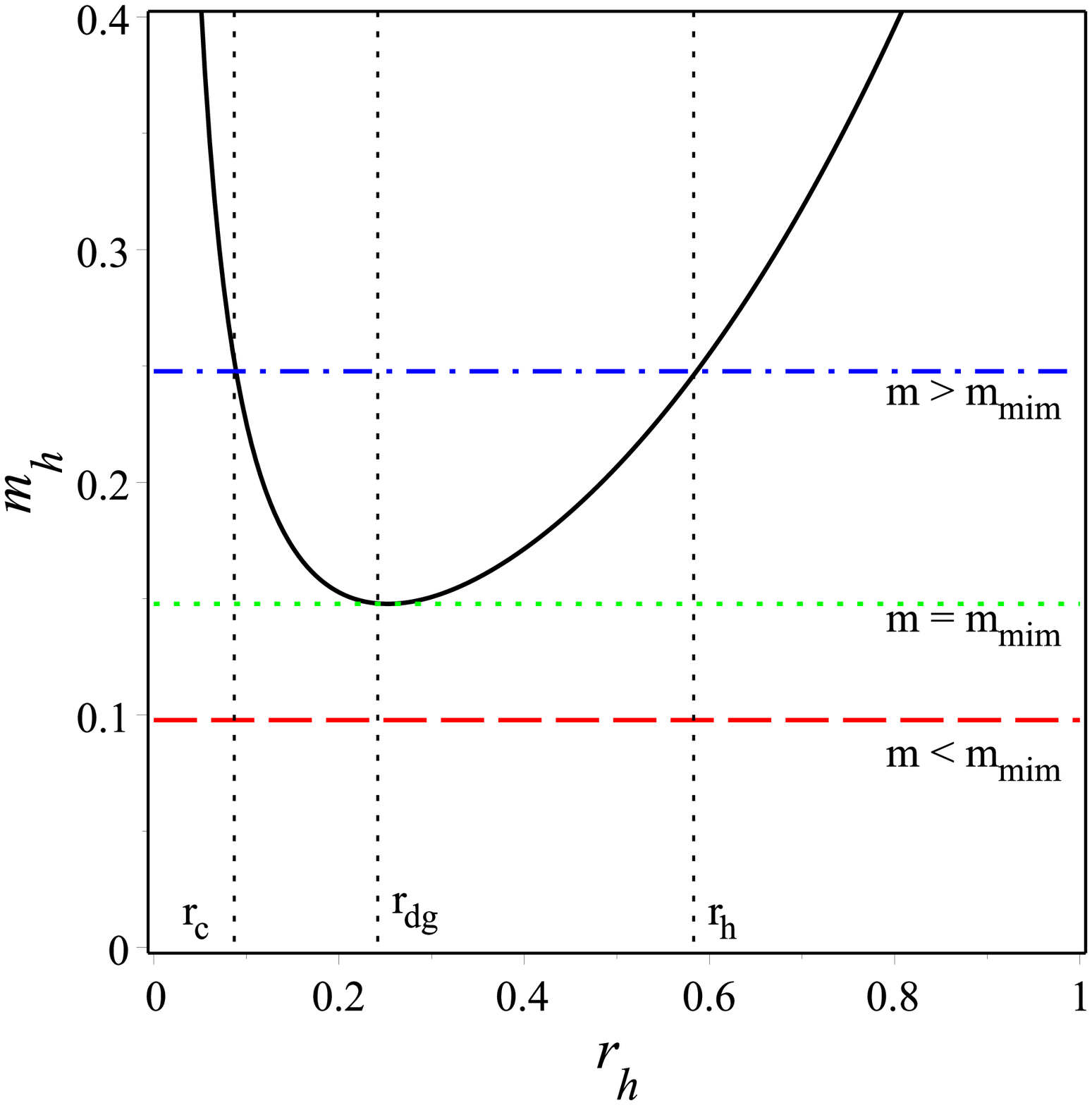}}
\subfigure[~The horizon Bekenstein-Hawking entropy]{\label{fig:2c}\includegraphics[scale=0.3]{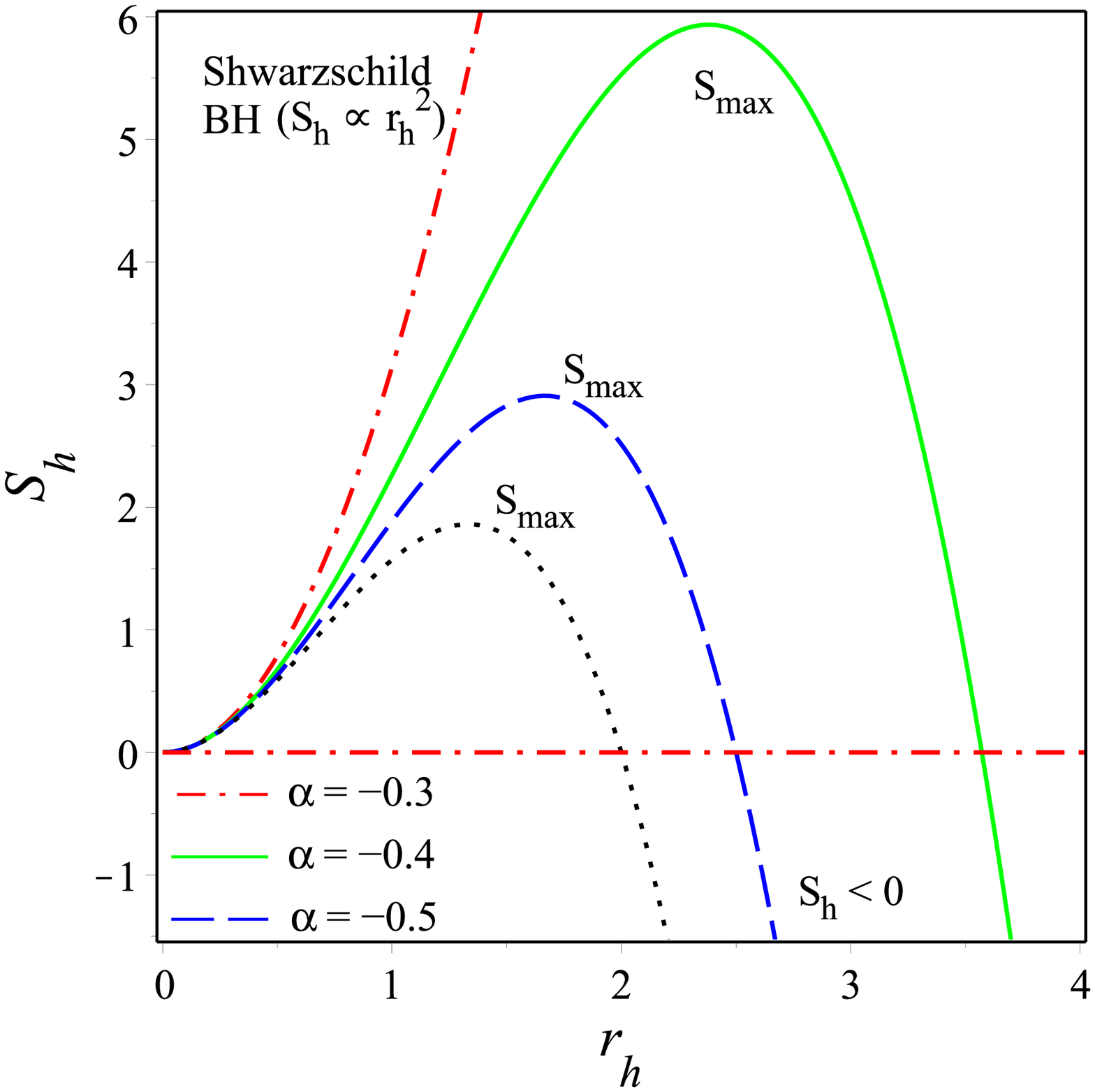}}\\
\subfigure[~The horizon Hawking Temperature]{\label{fig:2d}\includegraphics[scale=0.3]{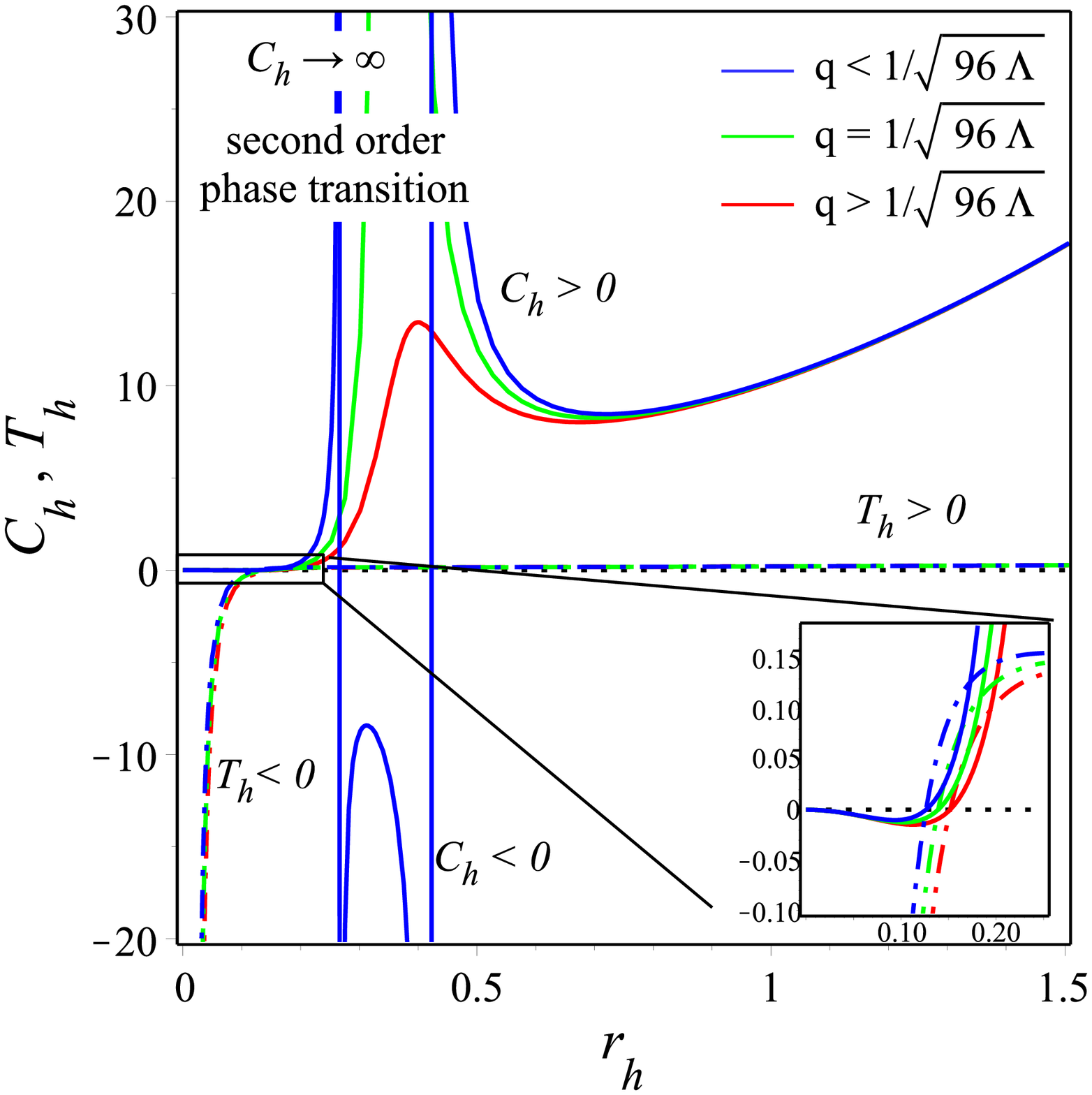}}
\subfigure[~The horizon quasi-local energy]{\label{fig:2e}\includegraphics[scale=0.3]{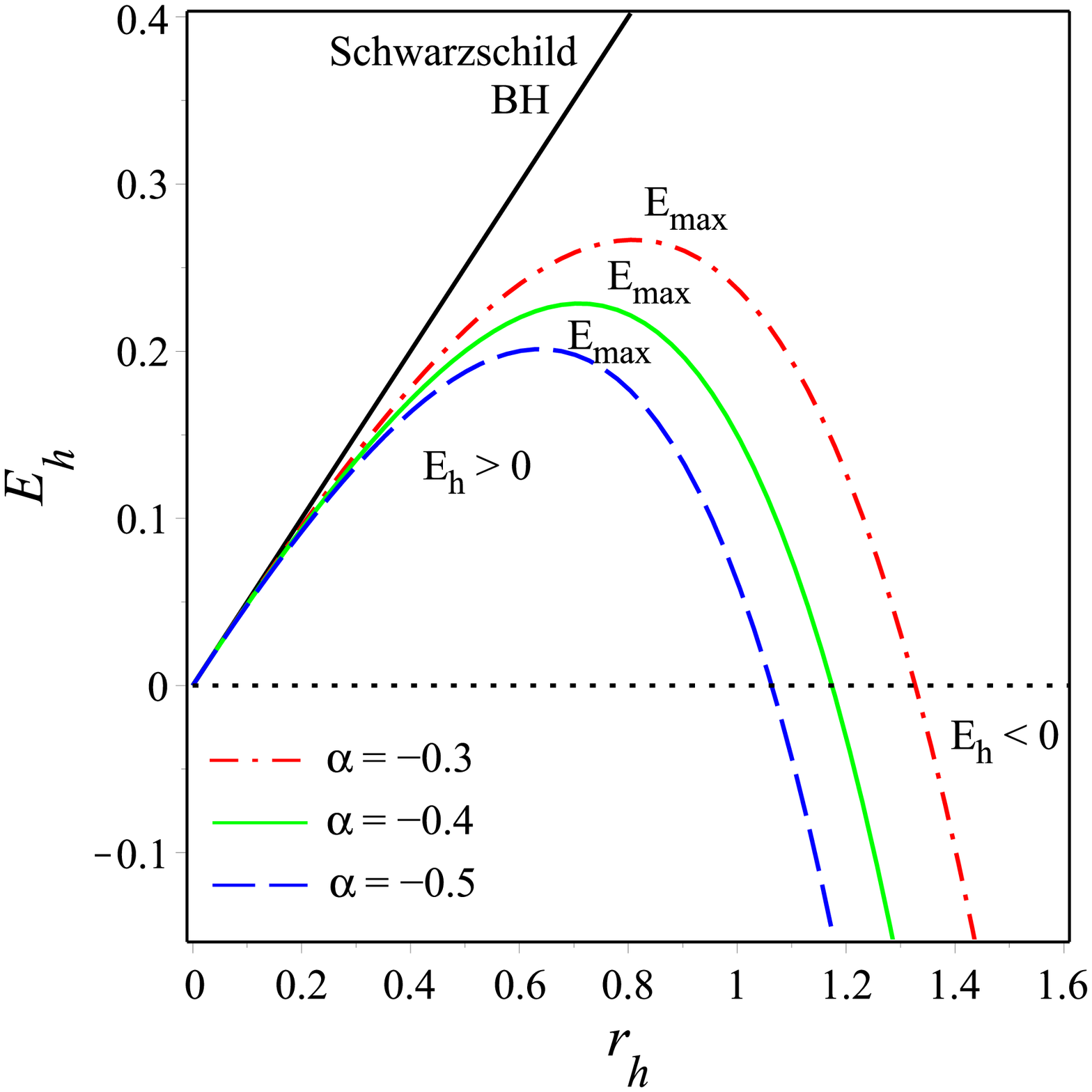}}
\subfigure[~The horizon Gibbs free energy]{\label{fig:2f}\includegraphics[scale=0.3]{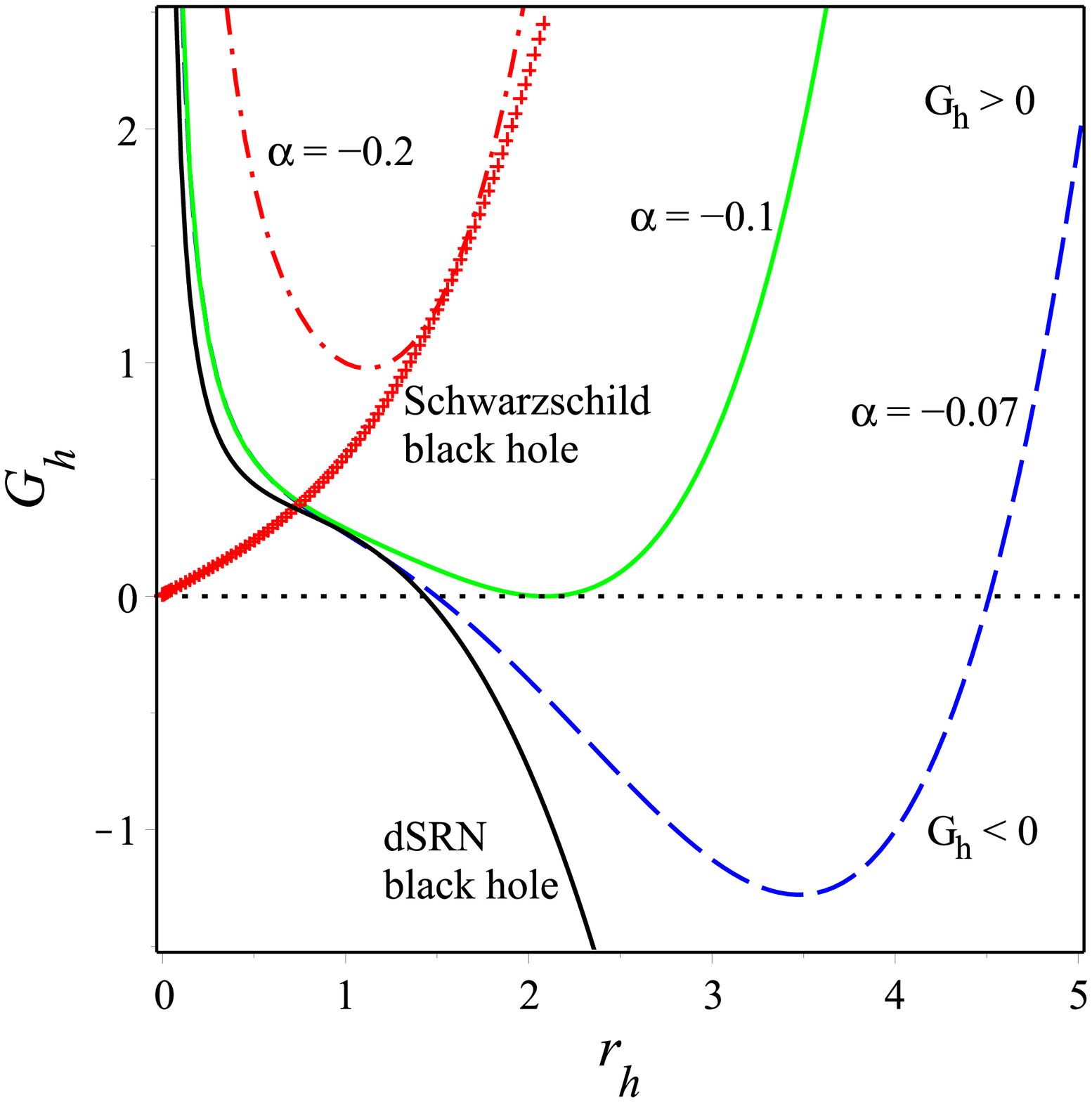}}
\caption{Schematic plots of thermodynamical quantities of the black hole solution (\ref{sol2a}) in Jordan frame: \subref{fig:2a} Typical behavior of the metric function $N(r)$ given by (\ref{sol2a}); \subref{fig:2b} the horizon mass--radius relation (\ref{hor-mass-rad2a}) with minimum mass given by (\ref{m_min2a}), it shows that, for $\Lambda>0$ case, the solution admits at most two horizons ($r_c$, $r_h$) when $m_h>m_{min}$, one degenerate horizon ($r_h=r_c=r_{dg}$) when $m_h=m_{min}$ and otherwise it provides a naked singularity; \subref{fig:2c} typical behavior of the horizon entropy, (\ref{BH-entropy2a}), which shows that $S_h$ does not always increases with $r_h$, rather it reaches a maximum $S_{max}$ at $r_h=\frac{2}{3|\alpha|}$ and then decreases, also it could have negative values when $r_h>1/|\alpha|$; \subref{fig:2d} typical behavior of the horizon temperature, (\ref{H-temp2a}), and the heat capacity, (\ref{heat-cap2a}), which shows that both vanish at $r_{dg}$, \eqref{deg_horz_2a}, whereas $T_h(r_h<r_{dg})<0$ and $T_h(r_h>r_{dg})>0$, for a critical charge $q=q_c=1/\sqrt{96\Lambda}$ the black hole admits a second order phase transition as $C_h$ diverges, for $q<q_c$ it acquires double second order phase transition, while for $0<q<q_c$ there is no such phase transition; \subref{fig:2e} typical behavior of the horizon quasi-local energy, (\ref{loc-eng2a}), which shows that $E_h$ could have negative values; \subref{fig:2f} typical behavior of the horizon Gibbs free energy, (\ref{G-eng2a}), which shows that $G_h$ could have negative values for some values of $\alpha$. We take $\Lambda=1$.}
\label{Fig:2}
\end{figure}

For the spherically symmetric black hole line element (\ref{line_element_1}), we study the $f(R)$ field equations (\ref{J-field-eqn1}) and (\ref{J-field-eqn2}). Here we introduce a new $f(R)$ theory,
\begin{equation}\label{fR2}
f(R)=R+2\alpha\sqrt{R +8\Lambda}\, ,
\end{equation}
where $\alpha$ is a nonzero dimensionful parameter.    We note that the above $f(R)$ gravity is similar to that has been studied in Ref.~\cite{Nashed:2019tuk}. Hence, we write the Jordan frame vacuum solution in presence of electromagnetism as follows,
\begin{equation} \label{sol2a}
N(r)=\frac{1}{2}-\frac{2m}{r}+\frac{2 \Lambda r^2}{3}+\frac{q^2}{r^2}\, ,
\end{equation}
where the mass of the black hole $m=-\frac{1}{6\alpha}$ which restricts $\alpha$ to have negative values. We take the cosmological constant $\Lambda>0$. Then the solution (\ref{sol2a}) produces dSRN black hole. Unlike the previous work \cite{Nashed:2019tuk}, the charge $q$ of the present solution is not related to the model parameter $\alpha$. Since the $\alpha$-parameter cannot vanish, the solution has no analogy in the GR theory. Interestingly, the solution produces a non-constant Ricci scalar, $R=-8\Lambda+\frac{1}{r^2}$, as required for our study in this section. Using the stability condition $f_R=1+\alpha r>0$, for $0<r<\infty$, we  restrict $\alpha$ to be negative, this is in agreement with the previous result we just got from the mass of the black hole. Since only a tiny correction of GR should be expected in this model, we find $\alpha\to 0^-$ and therefore $r$ can be extended up to $r\to \infty$. We note that solution (\ref{sol1a}) differs from solution (\ref{sol2a}) by a coefficient term $\frac{1}{2}$, which has several non-trivial consequences, e.g. the Ricci scalar of the later solution  has a radial dependence, while the former  is not. Remarkably, there is no re-parametrization that is mapping these two solutions. Similar to the quadratic polynomial $f(R)$ gravity, for AdSRN, i.e $\Lambda<0$, there are three horizons with one cosmological horizon while for dSRN spacetime, i.e $\Lambda>0$, there are only two event horizons. In Figure \ref{Fig:2}\subref{fig:2a}, we plot the later case of solution (\ref{sol2a}) which shows that the black hole could have at most two horizons, Cauchy (inner) horizon $r_c$ and the event (outer) horizon $r_h$, in the case that its horizon mass--radius exceeds a minimum,
\begin{equation}\label{m_min2a}
    m_{min}=\frac{(\sqrt{1+32 q^2 \Lambda}+2)(2\sqrt{1+32 q^2 \Lambda}-2)^{^{1/2}}}{24\sqrt{\Lambda}}\, .
\end{equation}
However, these two horizons coincide forming one horizon, a degenerate one, denoted as $r_{dg}$, in the case that the black hole mass is $m=m_{min}$, so we get a Nariai black hole. In all other cases there are no horizons. Notably, when $m = m_{min}$, we determine the degenerate horizon of the Nariai black hole as,
\begin{equation}\label{deg_horz_2a}
r_{dg}=\sqrt{\frac{-1+(1+32q^2\Lambda)^{1/2}}{8\Lambda}}\, .
\end{equation}
By setting $N(r_h) = 0$, we determine the horizon mass-radius relation,
\begin{eqnarray} \label{hor-mass-rad2a}
{m_h}=\frac{6q^2+3r_h^2+4\Lambda r_h^4}{12 r_h}.
\end{eqnarray}
In Fig. \ref{Fig:2}\subref{fig:2b}, we show the relation between the number of horizons and the horizon mass-radius relation. The Nariai black hole is located at $\partial m_h/\partial r_h=0$, which gives $r_{dg}$ as previously obtained in (\ref{deg_horz_2a}).
Next, we study the Bekenstein--Hawking entropy (\ref{BH-entropy}) for the solution (\ref{sol2a}), so we have,
\begin{eqnarray} \label{BH-entropy2a}
{S_h}=\frac{1}{4}A f_R=\pi  r_h^2 \left(1+\alpha r_h \right)\, .
\end{eqnarray}
As it is clear, the entropy is non-trivially modified, since $f_R(r_h)$ is not a constant. We note that the entropy is no longer proportional to the horizon area. In Fig. \ref{Fig:2}\subref{fig:2c}, we plot three possible behaviors of the Bekenstein--Hawking entropy, for different values of the parameter $\alpha$. The entropy increases monotonically to a maximum value $S_{max}=\frac{4\pi}{27\alpha^2}$ at $r_h=\frac{2}{3|\alpha|}$, then it decreases to zero at $\frac{2}{3|\alpha|}<r_h\leq \frac{1}{|\alpha|}$ and becomes negative for $r_h>\frac{1}{|\alpha|}$ as seen in Fig. \ref{Fig:2}\subref{fig:2c}. Remarkably, the entropy has a maximum value, which is not usual in the general relativistic black hole physics.

The horizon--temperature relation (\ref{Hawking-temp}) of the black hole solution (\ref{sol2a}) reads,
\begin{eqnarray} \label{H-temp2a}
{T_h}=\frac{r_h^2+4 \Lambda r_h^4-2q^2}{8\pi r_h^3}\, .
\end{eqnarray}
In addition, by substituting Eqs. (\ref{hor-mass-rad2a}) and (\ref{H-temp2a}) in (\ref{heat-capacity}), we calculate the heat capacity, which is,
\begin{equation} \label{heat-cap2a}
C_h=2\pi r_h^2\left[\frac{-2q^2+(1+4\Lambda r_h^2)r_h^2}{\;\;6q^2-(1-4\Lambda r_h^2)r_h^2}\right]\, ,
\end{equation}
We note that the behavior of the heat capacity is related to a critical charge $q_c=1/\sqrt{96\Lambda}$. For $q>q_c$, the heat capacity however diverges at,
$$r_{tr\pm}=\sqrt{\frac{1\pm(1-96\Lambda q^2)^{1/2}}{8\Lambda}}\, ,$$
which indicates a double second order phase transitions at these critical values $r_{tr\pm}$. In addition, the black hole is thermally unstable as the heat capacity acquires negative values when $r_{tr-}<r_h<r_{tr+}$, and it is thermally stable as the heat capacity becomes positive for $r_h>r_{tr+}$. For $q=q_c$, the heat capacity diverges once at $r_{h}=1/\sqrt{8\Lambda}$. For $0<q<q_c$, the heat capacity does not diverge at finite $r_h$ values. In Fig. \ref{Fig:2}\subref{fig:2d}, we plot the horizon--temperature relation (\ref{H-temp2a}) and the heat capacity (\ref{heat-cap2a}), showing their behaviors according to the choice of the model parameters. For the $\Lambda>0$ case, which we are focusing on, it is easy to show that the temperature and the heat capacity vanish on the degenerate horizon (\ref{deg_horz_2a}). In general, Hawking temperature and the heat capacity are negative where $r_h< r_{dg}$.

For the $f(R)$ gravity (\ref{fR2}), the quasi-local energy (\ref{local-energy}) becomes,
\begin{eqnarray} \label{loc-eng2a}
{E_h}=\frac{r_h}{2}\left[1+\alpha \, r \left(\frac{3}{4}+\Lambda r_h^2\right)\right]\, .
\end{eqnarray}
As clear the deviation from quasi-local energy of Schwarzschild black hole, $E_h=r_h/2$, is characterized by the model parameter $\alpha$. In Figure \ref{Fig:2}\subref{fig:2e}, we plot the horizon local energy showing possible positive and negative regions.

Finally, let us evaluate the Gibbs free energy of the solution (\ref{sol2a}), by substituting Eqs. (\ref{hor-mass-rad2a}), (\ref{BH-entropy2a}) and (\ref{H-temp2a}) in (\ref{Gibbs-energy}), which gives,
\begin{eqnarray} \label{G-eng2a}
{G_h}=\frac{1}{24\,r}\left[18q^2+3r^2-4\Lambda r^4 +3\alpha r \left(2q^2-r^2-4\Lambda r^4\right)\right]\, .
\end{eqnarray}
In Fig. \ref{Fig:2}\subref{fig:2f}, we plot the horizon Gibbs energy--radius relation, for small values of $|\alpha|$ as it is apparent from the plots, the black hole could become unstable for some values of $r_h$ wherever the Gibbs energy becomes negative. For larger values of $|\alpha|$, the black hole is stable as Gibbs energy $G_h>0$ and interpolates between dSRN and dS Schwarzschild black hole solutions.

In the next subsection, in order to accomplish our investigation of the (non)equivalence of Jordan and Einstein frames, we perform the same analysis of the above mentioned thermodynamical quantities--but--in Einstein frame.

%%%%%%%%%%%%%%%%%%%%%%%%%%%%%%%%%%%%%%%%%%%%%%%%%%%%%%%%%%%%%%%%%%%%%%%%%%%%%%%%%%%h3

\subsection{Thermodynamics of Black Holes in the Einstein frame}\label{S4.2}

\begin{figure}
\centering
\subfigure[~Possible two horizons]{\label{fig:3a}\includegraphics[scale=0.3]{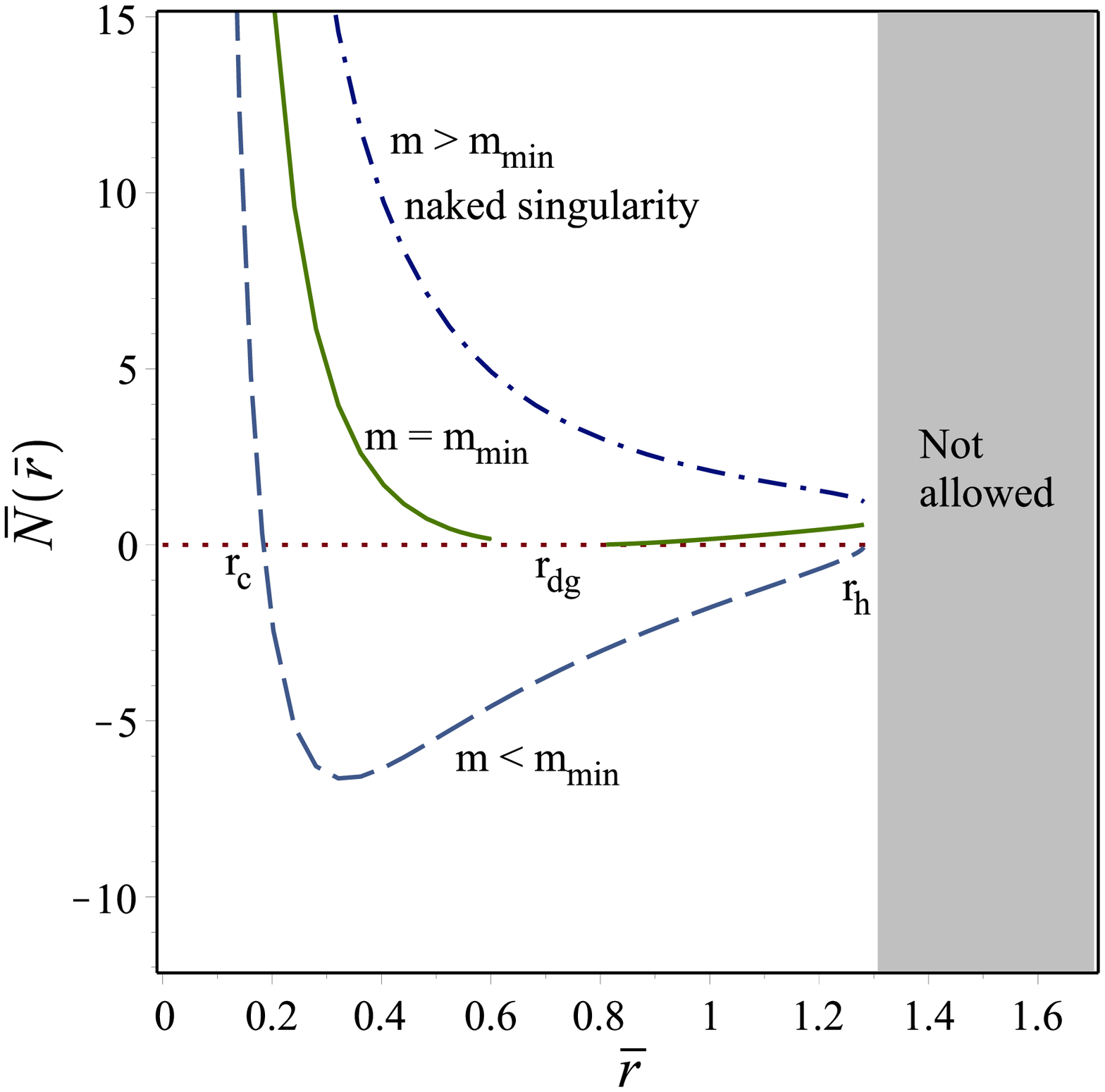}}
\subfigure[~The horizon mass-radius]{\label{fig:3b}\includegraphics[scale=0.3]{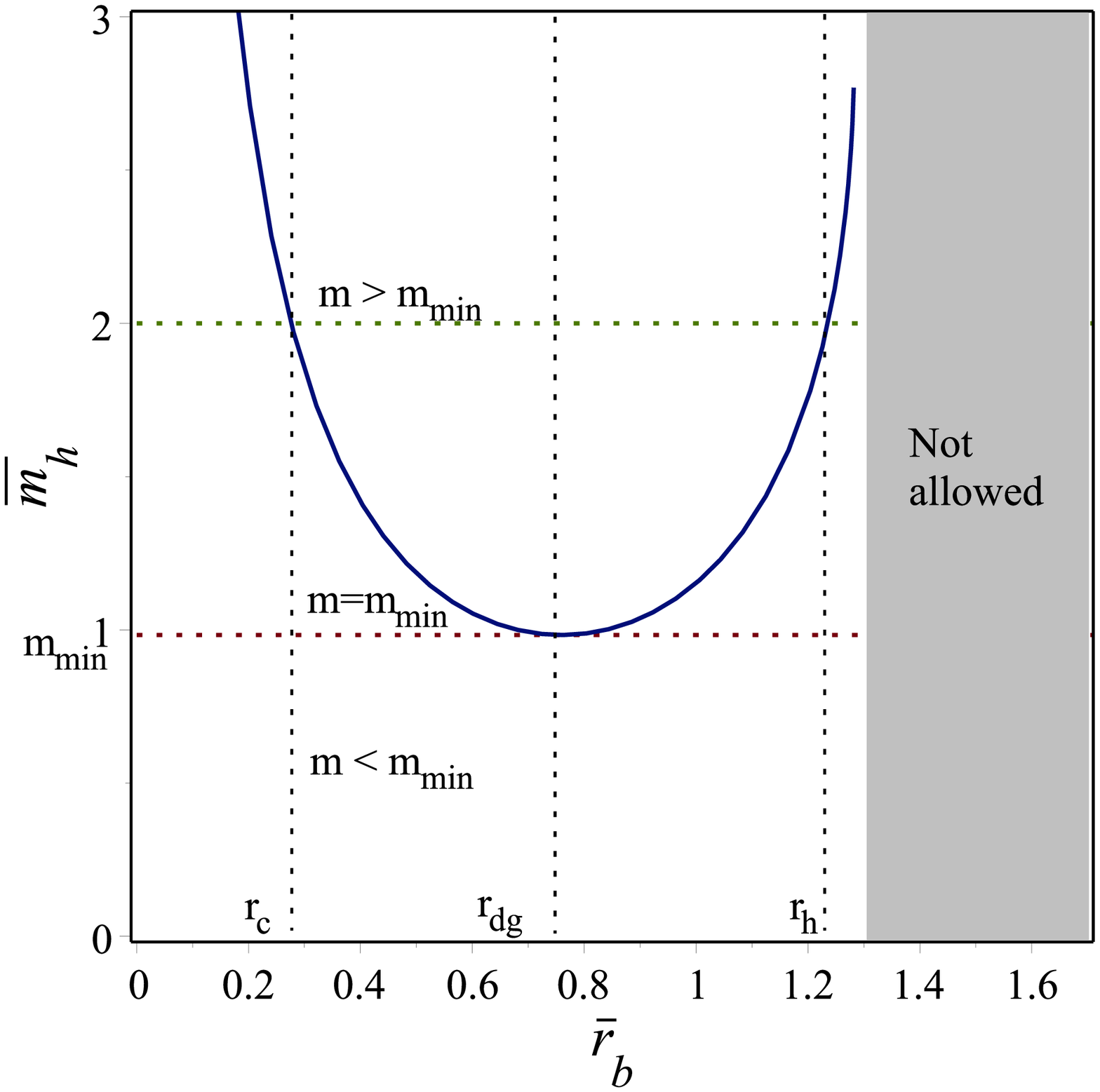}}
\subfigure[~The horizon Bekenstein-Hawking entropy]{\label{fig:3c}\includegraphics[scale=0.3]{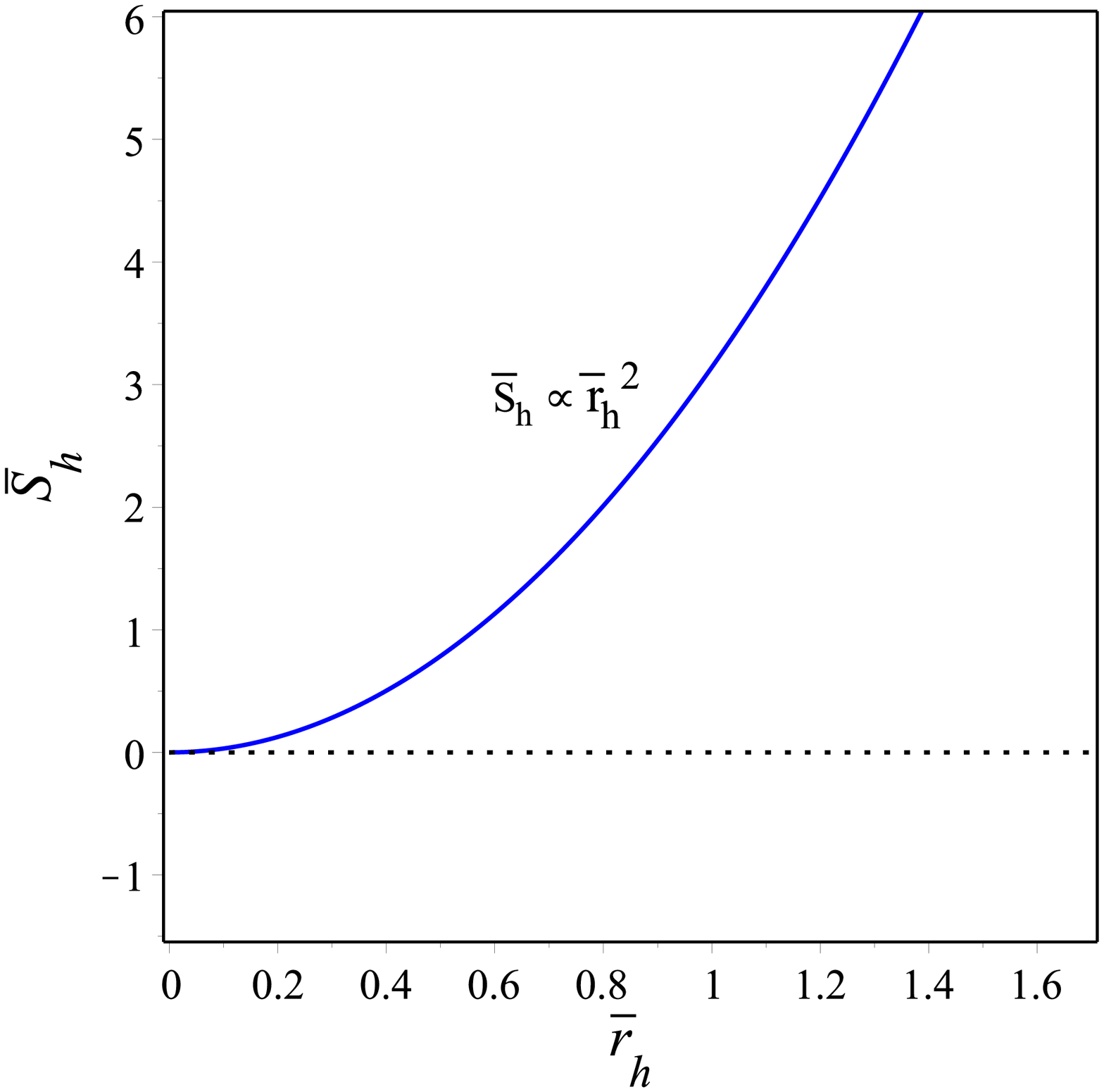}}\\
\subfigure[~The horizon Hawking Temperature]{\label{fig:3d}\includegraphics[scale=0.3]{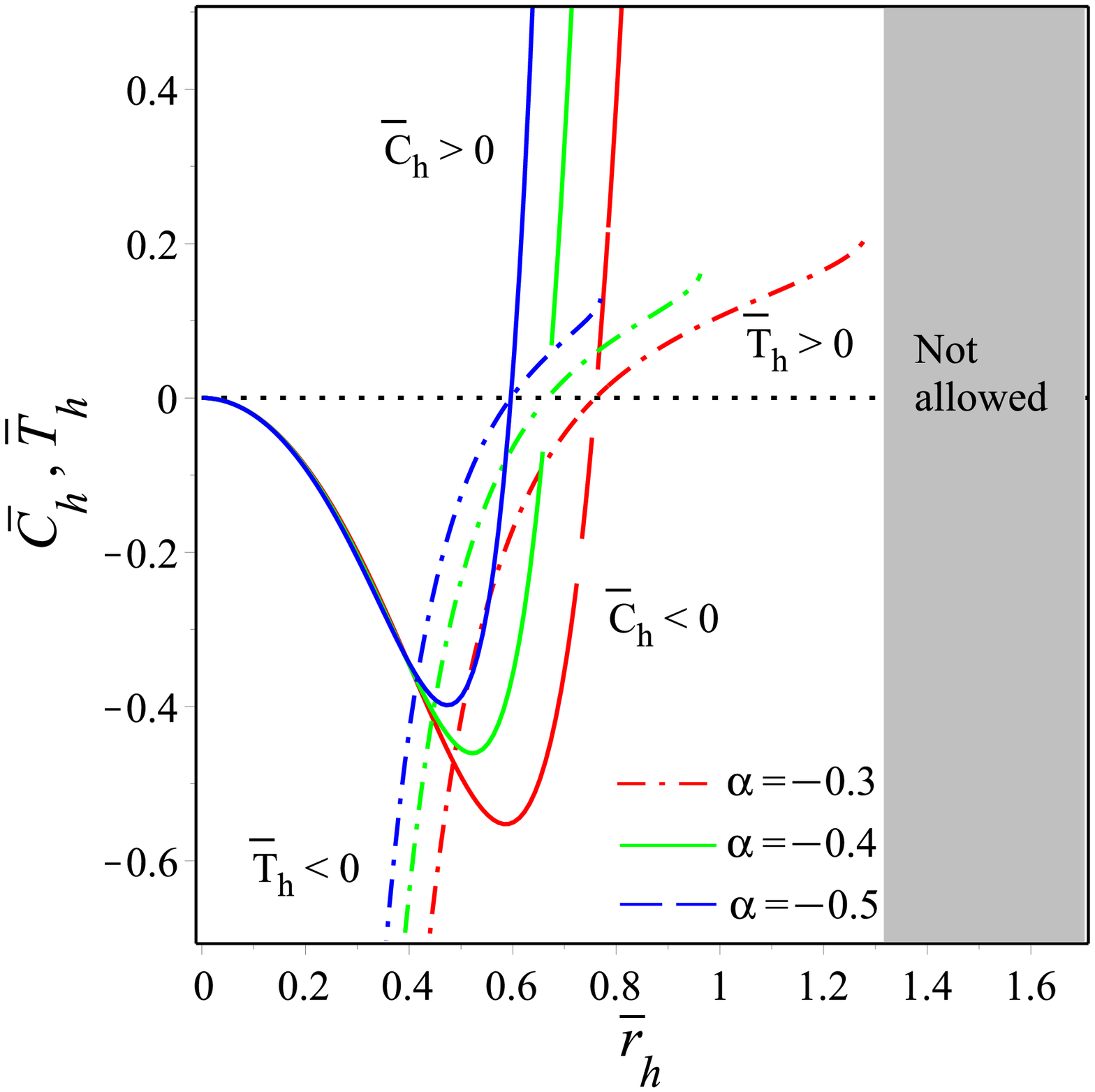}}
\subfigure[~The horizon quasi-local energy]{\label{fig:3e}\includegraphics[scale=0.3]{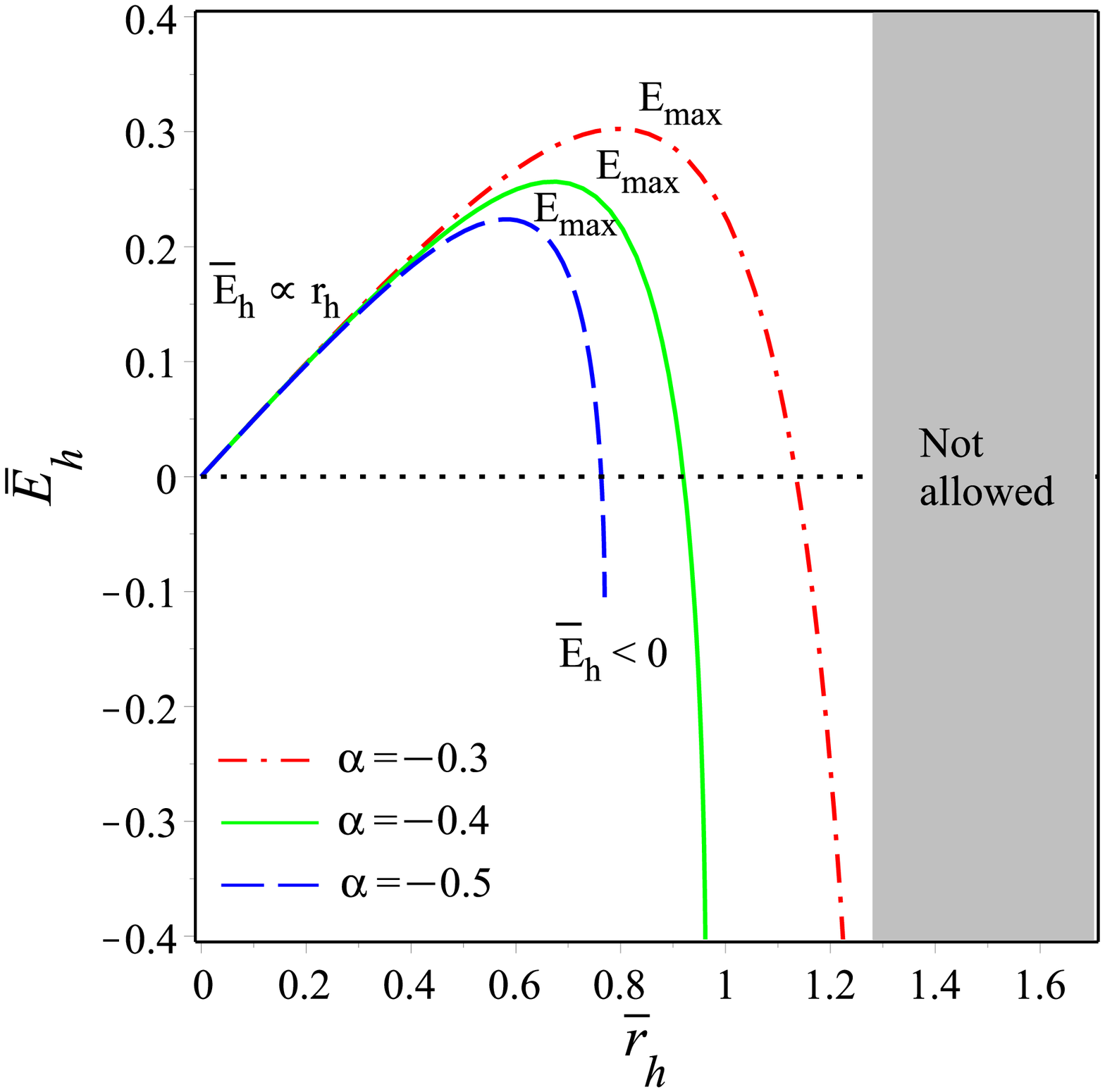}}
\subfigure[~The horizon Gibbs free energy]{\label{fig:3f}\includegraphics[scale=0.3]{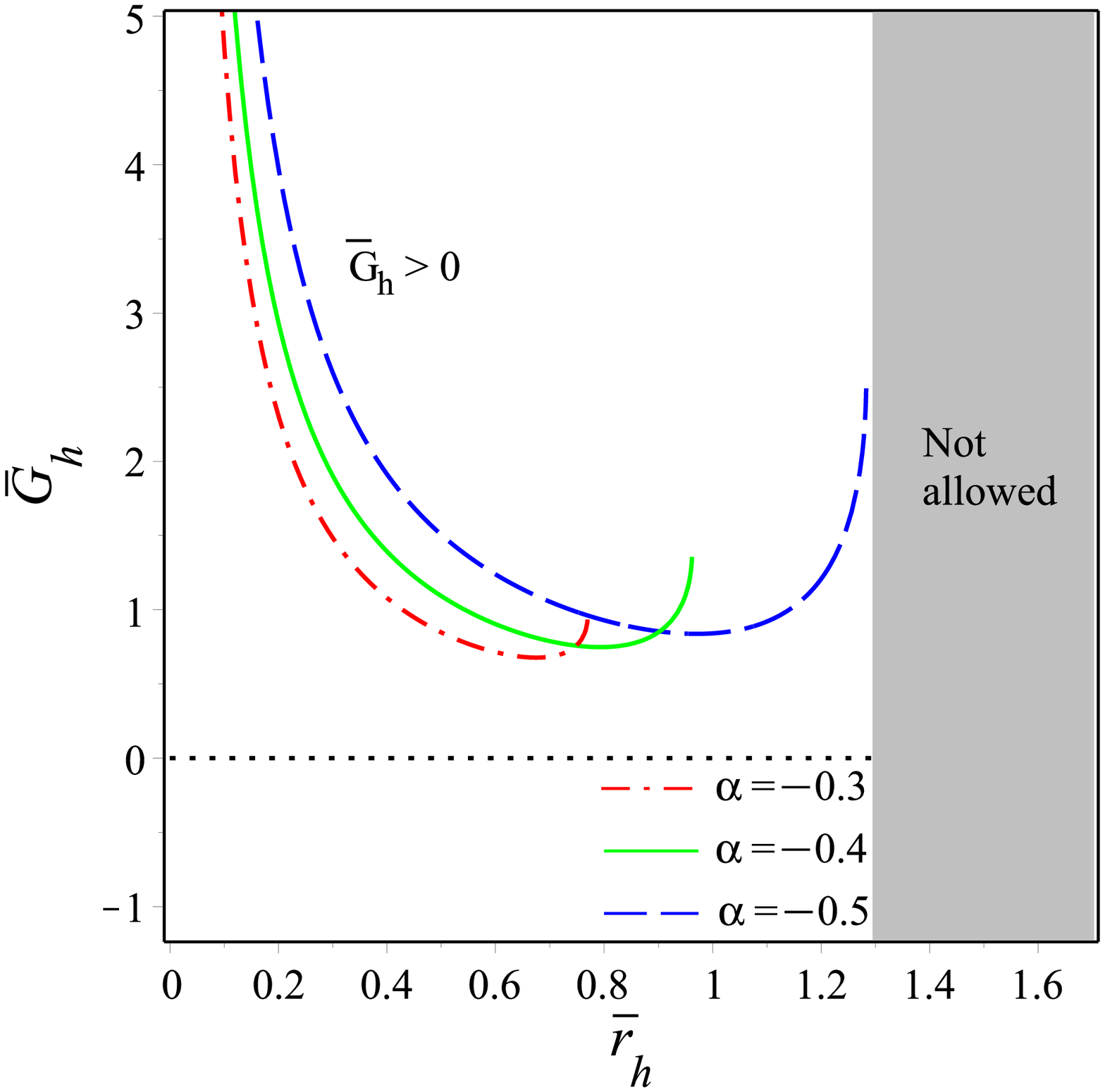}}
\caption{Schematic plots of thermodynamical quantities of the black hole solution (\ref{sol2b}) in Einstein frame: \subref{fig:3a} Typical behavior of the metric function $\bar{N}(\bar{r})$ given by (\ref{sol2b}); \subref{fig:3b} the horizon mass--radius relation \eqref{hor-mass-rad2b} with minimum mass given by (\ref{m_min2b}), it shows that, for $\Lambda>0$ case, the solution admits at most two horizons ($r_c$, $r_h$) when $m_h>m_{min}$, one degenerate horizon ($r_h=r_c=r_{dg}$) when $\bar{m}_h=m_{min}$ and otherwise it provides a naked singularity as long as the constraint on the radial coordinate $\bar{r}\leq \frac{2\sqrt{3}}{9\alpha}$ is fulfilled; \subref{fig:3c} typical behavior of the horizon entropy, (\ref{BH-entropy2b}), which shows that $S_h$ increases quadratically with $r_h$; \subref{fig:3d} typical behavior of the horizon temperature, (\ref{H-temp2b}), and the heat capacity, (\ref{heat-cap2b}), which shows that the black hole is not thermally stable (i.e. $\bar{C}_h<0$) whereas $\bar{r}_h<r_{dg}$; \subref{fig:3e} typical behavior of the horizon quasi-local energy, (\ref{loc-eng2b}), which shows that $E_h$ has maximum value and it could have negative values for some range of $r_{h}$; \subref{fig:3f} typical behavior of the horizon Gibbs free energy, (\ref{G-eng2b}), which shows that $G_h$ is always positive. As in Fig. \ref{Fig:2}, we take $\Lambda=1$. The comparison with the corresponding quantities in Jordan frame in Fig. \ref{Fig:2} shows clearly that the two frames are not equivalent from thermodynamics point of view.}
\label{Fig:3}
\end{figure}

In this subsection, we perform the analysis of the previous section in the Einstein frame, and we directly compare the results obtained in the Jordan and Einstein frame. For that purpose, we apply the conformal transformation (\ref{conf-trans}) to the spacetime metric (\ref{line_element_1}), that is $d\bar{s}_E^2=\Omega^2 ds_J^2$, where the conformal factor of the $f(R)$ gravity (\ref{fR2}) is given by,
\begin{equation}\label{conf-trans2}
\Omega^2=f_R=1+\alpha r, \, (0<r<1/|\alpha|).
\end{equation}
Thus, we write the Einstein frame metric,
\begin{eqnarray}\label{line_element_2E}
\nonumber    d\bar{s}_E^2&=& \Omega^2\left[-N(r) dt^2+\frac{dr^2}{N(r)}+r^2 \left(d\theta^2+\sin^2 \theta d\varphi^2\right)\right],\\
    &=&-\bar{N}(\bar{r}) d\bar{t}^2+\frac{d\bar{r}^2}{\bar{K}(\bar{r})}+\bar{r}^2 \left(d\theta^2+\sin^2 \theta d\varphi^2\right).
\end{eqnarray}
Here we take the coefficient of the solid angle to define the new radial coordinate, so we identify $\bar{r}=\Omega r$ as the radial coordinate in Einstein frame; consequently we define $\bar{N}(\bar{r})=\Omega^2 N(r)$, $\bar{K}(\bar{r})=\frac{N}{\Omega^2 \Sigma^2}$ and $\Sigma=\frac{dr}{d\bar{r}}$. We note that $\bar{N}(\bar{r})\neq\frac{1}{\bar{K}(\bar{r})}$. In addition, the radial coordinate transforms accordingly as,
\begin{equation}\label{conf-rad}
r\to \bar{r}=\sqrt{1+ \alpha r }\;\;r.
\end{equation}\
Since all physical quantities should be written in terms of $\bar{r}$, we write the inverse transformation,
\begin{equation}\label{inv-r-trans}
r=\frac{1}{\alpha}\left[-1+\left(\frac{1}{6}\chi^{1/3}+2\chi^{-1/3}\right)^2\right],
\end{equation}
where $\chi=108\alpha \bar{r}+12\sqrt{81\alpha^2 \bar{r}^{2}-12}$, which sets a constraint on the radial coordinate $\bar{r}\leq \frac{2\sqrt{3}}{9|\alpha|}$. This allows to write the potential (\ref{sol2a}) in the Einstein frame as,
\begin{eqnarray} \label{sol2b}
\nonumber\bar{N}(\bar{r})&=&\frac{-6912(\chi^{\frac{2}{3}}+\frac{1}{24}\chi^{\frac{4}{3}}+6)}{\alpha^2\chi^2(12\chi^{\frac{2}{3}}-\chi^{\frac{4}{3}}-144)^2} \left[\left( {-\frac {40}{27}}\,\Lambda-{\alpha}^{2} \right) {\chi}^{4/3}-
 \left( \frac{1}{48}\,{\alpha}^{2}+\frac{1}{8}\,{q}^{2}{\alpha}^{4}+\frac{1}{12}\,m{\alpha}^{
3}+{\frac {19}{972}}\,\Lambda \right) {\chi}^{\frac{8}{3}}\right.\\
\nonumber&&+ \left( {\frac {1}{
864}}\,{\alpha}^{2}+{\frac {1}{729}}\,\Lambda+{\frac {1}{144}}\,m{
\alpha}^{3} \right) {\chi}^{\frac{10}{3}}+{\frac {64}{9}}\,\Lambda\,{\chi}^{2/
3}+{\frac {1}{419904}}\,\Lambda\,{\chi}^{\frac{14}{3}}-{\frac {1}{20155392}}\,
\Lambda\,{\chi}^{\frac{16}{3}}+m{\chi}^{2}{\alpha}^{3}\\
&&\left.+ \left( \frac{1}{6}\,{\chi}^{2}
-{\frac {1}{20736}}\,{\chi}^{4} \right) {\alpha}^{2}- \left( {\frac {
64}{3}}+{\frac {5}{69984}}\,{\chi}^{4}-{\frac {16}{81}}\,{\chi}^{2}
 \right) \Lambda
\right].
\end{eqnarray}
We plot $\bar{N}(\bar{r})$ versus $\bar{r}$ in Fig. \ref{Fig:3}\subref{fig:3a}, which shows that the black hole is naked, if the mass is less than a minimum value, and particularly from,
\begin{equation}\label{m_min2b}
m_{min}=\sqrt{\frac{-1+96\Lambda q^2+(1+32\Lambda q^2)^{\frac{3}{2}}}{288\Lambda}}.
\end{equation}
Setting $\bar{N}(\bar{r})=0$ and solving for $m$, we get the horizon mass--radius relation,
\begin{eqnarray}\label{hor-mass-rad2b}
\nonumber \bar{m}_h(\bar{r}_h)&=&
\frac{1}{139968\alpha^3\chi^2}\left[\left( 20155392\,{\alpha}^{2}-29859840\,\Lambda \right) {\chi}^{4/3}+
\left( 419904\,{\alpha}^{2}-393984\,\Lambda+2519424\,{\alpha}^{4}{q}^{2}
\right) {\chi}^{8/3}\right.\\
\nonumber && -\left( 23328\,{\alpha}^{2}-27648\,\Lambda
\right) {\chi}^{10/3}+143327232\,\Lambda{\chi}^{2/3}+48\,\Lambda{\chi}^{14/3}-\Lambda{\chi}^{16/3}+972\,{\chi}^{2}
\left( {\chi}^{2}-3456 \right) {\alpha}^{2}\\
 && -\left.\left( 1440\,{\chi}^{4}+
429981696-3981312\,{\chi}^{2} \right) \Lambda\right]/(12\chi^{2/3}-\chi^{4/3}-144).
\end{eqnarray}
We plot the horizon mass--radius relation in Fig. \ref{Fig:3}\subref{fig:3b}, which shows that the black hole may have two horizons at most where $\bar{m}_h>m_{min}$ and it is naked singularity if $\bar{m}_h<m_{min}$. In addition, the plot shows the degenerate horizon $r_{dg}$ at which $m_h=m_{min}$, that is  when $\partial \bar{m}_h/\partial \bar{r}_h=0$ holds true.

Substituting Eq. (\ref{inv-r-trans}) into (\ref{BH-entropy2a}), we obtain that the black hole entropy in Einstein frame,
\begin{eqnarray} \label{BH-entropy2b}
{\bar{S}_h}=\pi  \, \bar{r}_h^2 =\frac{1}{4}\bar{A},
\end{eqnarray}
where $\bar{A}=4\pi \bar{r}^2_h$ is the surface area of the black hole event horizon. This matches the standard form of the black hole entropy in Einstein frame. Then, the black hole entropy $\bar{S}_h$ is positive and increases monotonically as it is apparent from Fig. \ref{Fig:3}\subref{fig:3c}. This is different from the corresponding behavior of the same quantity in the Jordan frame, namely Eq. (\ref{BH-entropy2a}). The entropy $S_h$ has a maximum value at a specific horizon radius $r_h$ and decreases at larger radii, which shows clearly that there is no thermodynamical equivalence. In general, one can observe that, for $f(R)$ gravity with non-constant Ricci scalar, the entropy in the Jordan frame ($S_h\propto r_h^2 f_R$) should include some terms with radial dependance different from the usual quadratic behavior of the Einstein frame. For sure if one transforms the area $\bar{A}\to \Omega^2 A=f_R A$, the Jordan frame entropy \eqref{BH-entropy2a} should be reproduced again. Similar conclusions have been obtained in the case of Brans Dicke theory \cite{Bhattacharya:2017pqc}. However, it is pointless to compare between the two frames while reexpressing the physical quantities in terms of the radial coordinates $r$ instead of $\bar{r}$. As we show clearly that the entropy in Jordan frame has been modified due to additional $\alpha$-term of the $f(R)$ gravity \eqref{fR2} as it should be. On the other hand, by applying the appropriate transformation, we obtain the standard picture $\bar{S}_h \propto \bar{r}^2_h$ of Einstein frame. We note that the inequivalence of Jordan and Einstein frames on the thermodynamic level has been discussed in the disformal $f(R)$ gravity \cite{Geng:2019wgd}.

Similarly, we substitute Eq. (\ref{inv-r-trans}) into (\ref{H-temp2a}) and (\ref{heat-cap2a}), thus the Hawking temperature and the entropy in Einstein frame associated with the black hole solution (\ref{sol2b}), respectively, read
\begin{eqnarray} \label{H-temp2b}
\nonumber {\bar{T}_h}=
\frac{-384}{\alpha \pi \chi^{2/3}}&&\left[\left( 243\,{\alpha}^{2}+1080\,\Lambda \right) \chi^{4/3}+
\left( \frac{81}{16}\,{\alpha}^{2}+\frac{57}{4}\,\Lambda-\frac{243}{8}\,q^2 {\alpha}^{4} \right) \chi^{8/3}\right.\\
\nonumber&& -\left( \frac{9}{32}\,{\alpha}^{2}+ \Lambda \right) \chi^{10/3}-5184\,\Lambda\chi^{2/3}-\frac{1}{576}\Lambda\chi^{14/3}+\frac{1}{27648}\,\Lambda\chi^{16/3}\\
&&\left.+ \left( \frac{5}{96}\,\chi^{4}-144\,\chi^{2}+15552
 \right) \Lambda+\frac {3}{256}\,\chi^{2} \left( \chi^{2}-
3456 \right) {\alpha}^{2}
\right]/(12 \chi^{2/3}-\chi^{4/3}-144)^3 ,
\end{eqnarray}
and
\begin{eqnarray} \label{heat-cap2b}
\nonumber\bar{C}_h&=&\frac{2\pi}{9\alpha^2 \chi^{4/3}} \left(\chi^{2/3}-\frac{1}{12}\chi^{4/3}-12\right)^2
\left[ \left( 243\,{\alpha}^{2}+1080\,\Lambda \right) \chi^{4/3}+
 \left( \frac{81}{16}\,{\alpha}^{2}+\frac{57}{4}\,\Lambda-\frac{243}{8}\,q^2 {\alpha}^{4} \right) \chi^{8/3}\right.\\
\nonumber &&- \left( \frac{9}{32}\,{\alpha}^{2}+\Lambda \right) \chi^{10/
3}-5184\,\Lambda\chi^{2/3}-\frac{1}{576}\Lambda\chi^{14/3}+\frac{1}{27648}\,\Lambda\chi^
{16/3}+ \left( \frac{5}{96}\,\chi^{4}-144\,\chi^{2}+15552
 \right) \Lambda\\
\nonumber&&\left.+{\frac {3}{256}}\,\chi^{2} \left( \chi^{2}-
3456 \right) {\alpha}^{2}\right]/\left[\left( \frac{57}{4}\,\Lambda-\frac{81}{16}\,{\alpha}^{2}+\frac{729}{8}\,q^2{\alpha}^{4} \right) \chi^{8/3}- \left( 243\,{\alpha}^{2}-1080\,\Lambda \right) \chi^{4/3}\right.\\
\nonumber&&+ \left( \frac{9}{32}\,{\alpha}^{2}-\Lambda \right) \chi^{10/3}-
5184\,\Lambda\chi^{2/3}-\frac{1}{576}\Lambda\chi^{14/3}+\frac{1}{27648}\,\Lambda\chi^{16
/3}+ \left( \frac{5}{96}\,\chi^{4}+144\,\chi^{2}-15552
 \right) \Lambda\\
&&\left.-{\frac {3}{256}}\,\chi^{2} \left( \chi^{2}-
3456 \right) {\alpha}^{2}\right].
\end{eqnarray}
We plot Hawking temperature and the heat capacity in Fig.  \ref{Fig:3}\subref{fig:3f}, which shows that both vanish at the event horizon. However, both have negative values as long as $\bar{r}_h<r_{dg}$ and therefore the black hole is not thermally stable. On the contrary, it is stable when $\bar{r}_h>r_{dg}$, since $\bar{C}_h>0$ and also $\bar{T}_h>0$.

Additionally, by substituting Eq. (\ref{inv-r-trans}) into (\ref{loc-eng2a}), the quasi-local energy of solution (\ref{sol2b}) is found to be,
\begin{eqnarray} \label{loc-eng2b}
\nonumber {\bar{E}_h} &=&
\frac{1}{3359232 \alpha^3\chi^{8/3}}\left[\left(23328\,{\alpha}^{2} -27648\,\Lambda \right) \chi^{10/3}+
 \left( 20155392\,{\alpha}^{2}+29859840\,\Lambda \right) \chi^{4/3}-
 \left( 139968\,{\alpha}^{2}- 393984\,\Lambda \right) \chi^{8/3}\right.\\
&&\left.-48\,\Lambda
\chi^{14/3}-\Lambda\chi^{16/3}-143327232\,\Lambda\chi^{2/3}+
 \left( 1440\,\chi^{4}-3981312\,\chi^{2}+429981696
 \right) \Lambda+972\,\chi^{2}{\alpha}^{2} \left( 3456+\chi^{2
} \right)
\right].\qquad
\end{eqnarray}
We plot the quasi-local energy in Fig. \ref{Fig:3}\subref{fig:3d}. Finally, by Substituting Eq. (\ref{inv-r-trans}) into (\ref{G-eng2a}), the free energy in the grand canonical ensemble is found to be,
\begin{eqnarray} \label{G-eng2b}
\nonumber {\bar{G}_h}&=&\frac{5}{486 \alpha^3 \chi^{\frac{8}{3}}(12\chi^{\frac{2}{3}}-\chi^{\frac{4}{3}}-144)}
\left[\left( -{\frac {1296}{5}}\,{\alpha}^{2}-{\frac {1152}{5}}\,\Lambda-{
\frac {11664}{5}}\,{q}^{2}{\alpha}^{4} \right) {\chi}^{\frac{10}{3}}+ \left( {
\frac {139968}{5}}\,{\alpha}^{2}+{\frac {684288}{5}}\,\Lambda \right)
{\chi}^{\frac{4}{3}}\right.\\
\nonumber &&+ \left( {\frac {11664}{5}}\,{\alpha}^{2}-{\frac {17496}{5
}}\,{q}^{2}{\alpha}^{4}+{\frac {12528}{5}}\,\Lambda \right) {\chi}^{\frac{8}{3}}- \left( {\frac {27}{40}}\,{\alpha}^{2}+\Lambda \right) {\chi}^{\frac{14}{3}}+ \left( {\frac {3}{320}}\,{\alpha}^{2}+{\frac {11}{240}}\,\Lambda
 \right) {\chi}^{\frac{16}{3}}-{\frac {2985984}{5}}\,\Lambda\,{\chi}^{\frac{2}{3}}\\
&&\left.+{
\frac {1}{34560}}\,\Lambda\,{\chi}^{\frac{20}{3}}-{\frac {243}{10}}
\,{\alpha}^{4}{\chi}^{4}{q}^{2}+{\frac {81}{5}}\,{\chi}^{2} \left( {
\chi}^{2}-864 \right) {\alpha}^{2}+{\frac {\Lambda}{720}}\, \left(
12528\,{\chi}^{4}-{\chi}^{6}+14929920\,{\chi}^{2}
+1289945088 \right)
\right].\qquad
\end{eqnarray}
The behaviors of the Gibbs energy of the black holes studied are presented in Fig. \ref{Fig:3}\subref{fig:3f}  for some values of the model parameters. As it can be seen in the figure, for the black hole solution (\ref{sol2b}) the Gibbs energy is always positive.

Therefore, the conclusion of this section is that for non-constant scalar curvature black holes, the results in the Jordan and Einstein frames are different. In the next section, we complete our investigation of the correspondence of Jordan and Einstein frames on the cosmologic scale as well.

%%%%%%%%%%%%%%%%%%%%%%%%%%%%%%%%%%%%%%%%%%%%%%%%%%%%%%%%%%%%%%%%%%%%%%%%%%%%%%%%%%%%%%%%%%%%%%%%%%%%%%h5

\section{Thermodynamics of the Power-law cosmology in $f(R)$ Gravity}\label{S5}

Similar to sections \ref{S3} and \ref{S4}, we discuss here the thermodynamics of cosmological solutions of an $f(R)$ gravity in both the Jordan and Einstein frames, focusing on those cosmologies with non-trivial scalar curvature. The aim is to investigate the validity of the correspondence of these solutions, in both frames, on the thermodynamics level. Unlike the previous sections, we begin here with the Einstein frame, then we discuss the correspondence in the Jordan frame.

%%%%%%%%%%%%%%%%%%%%%%%%%% Section 5.1 %%%%%%%%%%%%%%%%%%%%%%%%%%%%%
\subsection{Thermodynamics in the Einstein Frame}\label{S5.1}
%%%%%%%%%%%%%%%%%%%%%%%%%%%%%%%%%%%%%%%%%%%%%%%%%%%%%%%%%%%%%%%%%%%%

In this subsection, we consider the Fonarev solution, which describes a spherically symmetric and dynamically inhomogeneous background \cite{Fonarev:1994xq}. This solution is asymptotically an FLRW geometry, generated by a minimally coupled self-interacting scalar field $\phi$ with an exponential potential $V(\phi)$. In particular, for a vanishing mass parameter, the line element reduces to the spatially flat FLRW metric,
\begin{equation} \label{ds2}
d\bar{s}_E^2=-d\bar{t}^2+\bar{a}(\bar{t})^2\Big[dr^2+r^2\Big\{d\theta^2+\sin^2\theta d\varphi^2\Big\}\Big],
\end{equation}
where $\bar{a}(\bar{t})$ is the scale factor. In the following, we adopt the reduced Planck system of physical units, in which, $\hbar=G=c=k_B=1$. In the Einstein frame, the solution set of the field equations of the action (\ref{E-action1}) is given as \cite{Fonarev:1994xq, Faraoni:2017afs},
\begin{eqnarray}
\bar{a}(\bar{t})&=&(4a_1 \alpha_1^2 \bar{t})^{\frac{1}{4\alpha_1^2}},\label{sc}\\
\phi(\bar{t})&=&\frac{1}{\alpha_1\sqrt{2\kappa}}\ln(4a_1 \alpha_1^2 \bar{t}),\label{sc-field}\\
V(\phi)&=&\frac{a_1^2(3-4\alpha_1^2)}{\kappa}e^{-\alpha_1\sqrt{8\kappa} \phi},\label{pot}
\end{eqnarray}
where $\alpha$ and $a_1$ are two constants. By comparing with the standard power-law scale factor $a\propto t^{\frac{2}{3(1+w_\phi)}}$ with $a(t_0)=1$ at present time $t_0$, it is not difficult to show that $\alpha_1=\frac{1}{4}\sqrt{6(1+w_\phi)}$ and $a_1=\frac{1}{\bar{t}_0 \sqrt{6(1+w_\phi)}}$, which gives $a=\left(\frac{\bar{t}}{\bar{t}_0}\right)^{\frac{2}{3(1+w_\phi)}}$ where $w_{\phi}$ is the equation of state parameter. Also, the Hubble and the deceleration parameters are defined as follows,
\begin{equation} \label{ah}
\bar{H}=\frac{1}{\bar{a}}\frac{d\bar{a}}{d\bar{t}}=\frac{2}{3 (1+w_\phi) \bar{t}}\;, \qquad \qquad \bar{q}=-1-\frac{1}{\bar{H}^2}\frac{d\bar{H}}{d\bar{t}}=\frac{1}{2}(1+3w_\phi)\, .
\end{equation}
We note that the solution produces an accelerating expansion with $w_\phi < -\frac{1}{3}$. Thus, the Ricci scalar reads,
\begin{equation} \label{Rc}
\bar{R}=6(d\bar{H}/d\bar{t}+2\bar{H}^2)=\frac{4(1-3w_\phi)}{3(1+w_\phi)^2  \bar{t}^2}\, ,
\end{equation}
which is obviously a dynamical function. Next, we calculate some thermodynamical quantities associated of the solutions (\ref{sc})--(\ref{pot}), and in particular, the Hawking Temperature $T=\frac{\tilde{\kappa}}{2\pi}$ where $\tilde{\kappa}$ denotes the surface gravity of the cosmological horizon and the entropy $S=\mathcal{A}/4$ where $\mathcal{A}$ denotes the area of the cosmological horizon. In practice, we deal with three cosmological horizons, namely the apparent, the Hubble and the event horizon. However, we follow the argument that the apparent horizon, for a dynamical spacetime, is the causal one associated with the surface gravity and the gravitational entropy \cite{Hayward:1997jp,Hayward:1998ee,Bak:1999hd,Cai:2005ra}.  For this purpose, we write the apparent horizon $\mathcal{R}_{AH}$ of the Fonarev solution,
\begin{equation} \label{ah}
\bar{\mathcal{R}}_{AH}=\frac{1}{\bar{H}} =\frac{3}{2}(1+w_\phi) \bar{t},
\end{equation}
which coincides with the Hubble horizon in the flat FLRW case. Thus, the Hawking temperature $T_{AH}$ on the apparent horizon $\mathcal{R}_{AH}$ can be written as \cite{Cai:2005ra},
\begin{equation} \label{ah1}
\bar{T}_{AH}=\frac{-1}{2 \pi \bar{\mathcal{R}}_{AH}}\Big(1-\frac{d\bar{\mathcal{R}}_{AH}/d\bar{t}}{2\bar{H} \bar{\mathcal{R}}_{AH}}\Big)=-\frac{1-3w_\phi}{12 \pi (1+w_\phi) \bar{t}}\, .
\end{equation}
Using (\ref{ah}), we get the entropy,
\begin{equation} \label{ah4}
\bar{S}=\frac{1}{4}\mathcal{A}= \frac{9}{4}\pi (1+w_\phi)^2 \bar{t}^2,
\end{equation}
where the area of the apparent horizon is given as $\mathcal{A}=4 \pi \mathcal{R}_{AH}^2$. %Using Friedmann's equation, we write
Now let us consider the same theoretical framework in the Jordan frame, which is the subject of the next section. We shall mainly focus on quantities depending on the Hubble rate.

%%%%%%%%%%%%%%%%%%%%%%%%%%%%%%%%%%%%%%%%%%%%%%%%%%%%%%%%%%%%%%
\subsection{Thermodynamics in the Jordan frame}\label{S5.2}
%%%%%%%%%%%%%%%%%%%%%%%%%%%%%%%%%%%%%%%%%%%%%%%%%%%%%%%%%%%%%%

In order to calculate the thermodynamical quantities in the Jordan frame, we determine the $f(R)$ gravity which corresponds to the scalar potential (\ref{pot}). We define $\alpha=\sqrt{12}\alpha_1$ and $V_0=2a_1^2(3-4\alpha_1^2)$, hence the potential (\ref{pot}) reads,
$$V(\phi)=\frac{V_0}{2\kappa}e^{-\sqrt{\frac{2\kappa}{3}}\alpha\phi}\, .$$
Taking into account the relation (\ref{scalar-field}) and by substituting into (\ref{pot-fR}), we obtain,
\begin{equation}\label{const_fR3}
    V_0 f_R^{2-\alpha}-R f_R+f(R)=0\, .
\end{equation}
The solution of the above differential equation gives,
\begin{equation}\label{fR3}
    f(R)=(-1)^{n}(1-\alpha)(\alpha-2)^{-n} V_0^{\frac{1}{\alpha-1}}R^n\, ,
\end{equation}
where $n=\frac{\alpha-2}{\alpha-1}$. In this case, with the help of equation (\ref{sc-field}), we evaluate the conformal factor,
\begin{equation}\label{conf_fac3}
    \Omega(\bar{t})=(1/6)^{1/\alpha}\alpha^{2/\alpha}\left(\sqrt{\frac{6V_0}{9-\alpha^2}}\bar{t}\right)^{1/\alpha}\, .
\end{equation}
Now by conformally transforming the metric, $ds_J^2=\Omega^{-2}d\bar{s}_E^2$, we get the Jordan frame FLRW metric,
\begin{eqnarray}
ds_J^2&=&\Omega^{-2}\left[-d\bar{t}^2+\bar{a}(\bar{t})^2\left(dr^2+r^2\Big\{d\theta^2+\sin^2\theta d\varphi^2\Big\}\right)\right],  \label{ds2J} \\
&=&-dt^2+a(t)^2\left(dr^2+r^2\Big\{d\theta^2+\sin^2\theta d\varphi^2\Big\}\right), \label{ds2J2}
\end{eqnarray}
where,
$$dt=\Omega^{-1}d\bar{t},\quad a(t)=\Omega^{-1}\bar{a}(\bar{t}).$$
Thus, the Hubble and the deceleration parameters of the line-element (\ref{ds2J}), in the Jordan frame, are given by,
\begin{equation} \label{ahr11}
H=\frac{3+\alpha}{\alpha(1+\alpha)t},\qquad q=-1+\frac{\alpha(1+\alpha)}{3+\alpha}.
\end{equation}
It is obvious that these quantities have similar dynamical behavior to the ones corresponding to the Einstein frame. Hence the thermodynamical quantities, which depend on Hubble in the two frames, are similar, however this result might be model dependent. Indeed, as was shown in Ref. \cite{Bahamonde:2017kbs}, certain types of singularities have different form in the two frames. Thus our result is not a general result but it rather shows the accidental similarity of some quantities which are not conformal invariant in the two frames. From a mathematical point of view, the answer behind this similarity might be found in the dynamical systems corresponding to the Einstein and Jordan frame cosmological systems, and specifically on how the trajectories in the phase space are related via a conformal transformation. This is a longstanding challenging mathematical problem, which we need to address separately in a future work.

%%%%%%%%%%%%%%%%%%%%%%%%%%%%%%%%%%%%%%%%%%%%%%%%%%%%%%%%%%%%%%%%%%%%%%%%%%%%%%%%%%%%%%%%%%%%%%%%%
\section{Thermodynamics of UV $f(R)$ Gravity}\label{S6}
%%%%%%%%%%%%%%%%%%%%%%%%%%%%%%%%%%%%%%%%%%%%%%%%%%%%%%%%%%%%%%%%%%%%%%%%%%%%%%%%%%%%%%%%%%%%%%%%%
In this section we introduce a new $f(R)$ theory which is compatible with high energy scales at inflation. In Jordan frame, we expect a cosmology different from the power-law model due to the contribution of the higher order corrections of the adopted $f(R)$ gravity. Then, in Einstein frame, we derive the corresponding scalar field which generates the Jordan $f(R)$ gravity. We briefly discuss the scalar field potential pattern and its consequences on inflation. Finally, we compare the thermodynamical quantities in both Jordan and Einstein frames.
%%%%%%%%%%%%%%%%%%%%%%%%%% Section 6.1 %%%%%%%%%%%%%%%%%%%%%%%%%%%%%
\subsection{Thermodynamics in the Jordan Frame}\label{S6.1}
%%%%%%%%%%%%%%%%%%%%%%%%%%%%%%%%%%%%%%%%%%%%%%%%%%%%%%%%%%%%%%%%%%%%
Applying the $f(R)$ field equations to the FLRW spacetime \eqref{ds2J2} in Jordan frame, we write the modified Friedmann equations
\begin{eqnarray}
3H^2&=&\frac{\kappa}{f_R}\left(\rho_M+\rho_{R}\right),\label{fRinf1}\\
2\dot{H}+3H^2&=&-\frac{\kappa}{f_R}\left(p_M+p_R\right),\label{fRinf2}
\end{eqnarray}
where the $\rho_R$ and $p_R$ are the quantities other than Einstein tensor. Although they have geometrical origins too, it is useful to separate them on the right hand sides in order to identify and interpret the contribution of the modified gravity. These are given by
\begin{eqnarray}
\rho_R&=&-\frac{1}{2\kappa}\left( f - R f_R + 6H\dot{R} f_{RR}\right),\label{dens_curv}\\
p_R&=&\frac{1}{2\kappa}\left[f - R f_R + 2\left(\ddot{R}+2H\dot{R}\right)f_{RR}+2\dot{R}^2 f_{RRR} \right].\label{press_curv}
\end{eqnarray}
Now it is obvious to see that the GR theory is recovered by setting $f(R)=R$, where $\rho_R$ and $p_R$ vanish identical and \eqref{fRinf1} and \eqref{fRinf2} reproduce the standard Friedmann and Raychaudhuri equations.

We introduce a novel $f(R)$ gravity of the form
\begin{equation}
\label{UV_fRinf}
f(R)=R\,e^{\alpha R/R_i},
\end{equation}
where the constant $R_i$ denotes the value of the Ricci scalar at inflation and $\alpha$ is a dimensionless parameter. Obviously, the above form covers the ultraviolet (UV) gravity, since it may take the form $f(R)\simeq R+\alpha R^2/R_i+\frac{1}{2} \alpha^2 R^3/R_i^2+ \cdots$, ($\alpha R/R_i\lesssim 1$). So we expect that the $f(R)$ corrections to have an essential role at high energy scales, $R\simeq R_i$. In addition, the $f(R)$ gravity \eqref{UV_fRinf} fulfills the constraints $f_R>0$ and $f_{RR}>0$, which indicates that ghosts or gravitational instabilities are not expected. However, at low energy scale, $\alpha R/R_i\ll 1$, the theory reproduces the GR gravity, whereas $f(R)\simeq R$. Recalling that $R=6(2H^2+\dot{H})$, in absence of matter ($\rho_M=0,~p_M=0$), Eq.~(\ref{fRinf1}) and (\ref{fRinf2}) give
\begin{eqnarray}
\nonumber & & \left[12\alpha \left(6\alpha H^2+3\alpha \dot{H}+R_i\right)H \ddot{H}+6\alpha\left(24\alpha H^2-R_i\right)\dot{H}^2+6\alpha \left(48\alpha H^2+5 R_i\right)H^2\dot{H}\right.\\
&&\qquad\left.-\left(12\alpha H^2+R_i\right) R_i H^2\right] e^{\frac{6\alpha(2H^2+\dot{H})}{R_i}}=0\,,
\label{UVeq1}\\[5pt]
\nonumber & & \left[4\alpha R_i (6\alpha H^2+3\alpha \dot{H}+R_i)\dddot{H}+36\alpha^2 (R_i+4\alpha H^2+2\alpha \dot{H})\ddot{H}^2+28\alpha \left(\frac{144}{7}\alpha^2\dot{H}^2+\frac{93}{7}\alpha(R_i+\frac{96}{31}\alpha H^2)\dot{H}\right.\right.\\
\nonumber &&\left.+R_i(R_i+6\alpha H^2)\right)H \ddot{H}++48\alpha^2 (R_i+24\alpha H^2)\dot{H}^3+2\alpha(408 \alpha R_i H^2+7R_i+1152 \alpha^2 H^4)\dot{H}^2\\
&&\left. +\frac{2}{3}R_i (R_i+9\alpha H^2)(R_i +48 \alpha H^2)\dot{H}+R_i^2 H^2 (R_i-12 \alpha H^2)\right] e^{\frac{6\alpha(2H^2+\dot{H})}{R_i}} =0\,.\label{UVeq2}
\end{eqnarray}
During inflation the $\dot{H}^2$, $\dot{H}^3$, $\ddot{H}$ and $\dddot{H}$ terms in Eq.~(\ref{UVeq1}) and \eqref{UVeq2}
can be neglected relative to others. In addition, we note that the exponential term, $e^{6\alpha R/R_i}$, cannot be made to vanish. The above equations read
\begin{eqnarray}
6\alpha \left(48 \alpha H^2 +5 R_i\right) \dot{H} -12 \alpha R_i H^2+R_i^2&=&0,\\
2(R_i+57\alpha R_i H^2+432 \alpha^2 H^4)\dot{H}+3R_i^2 H^2 - 36 \alpha R_i H^4 &=&0.
\end{eqnarray}
We next obtain the solution
\begin{figure}
\begin{center}
\includegraphics[scale=0.4]{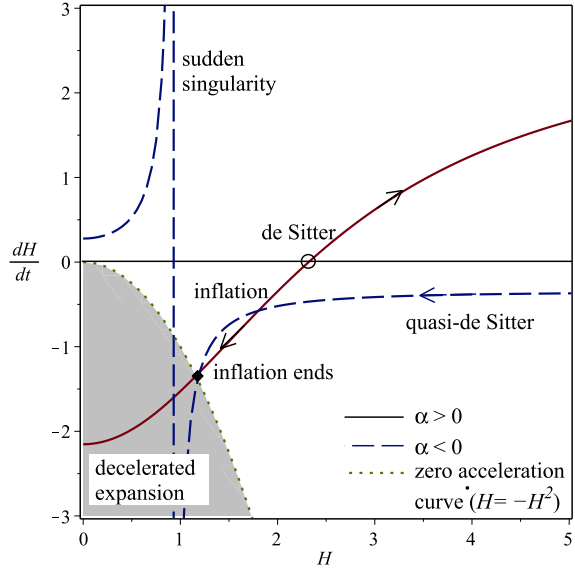}
\caption{($H$,~$\dot{H}$ ) phase space: For $\alpha>0$, the trajectory has a fixed point (de Sitter) at $H_f=\frac{1}{6}\sqrt{\frac{3R_i}{\alpha}}$. Since the system cannot reach this point in finite time, the universe has no initial singularity. Therefore, for $H<H_f$, it evolves in inflationary regime with $\dot{H}<0$ which can end the inflation epoch naturally as indicated by the intersection of the phase portrait with the zero acceleration curve, $\dot{H}=-H^2$. For $\alpha<0$, the universe begins with a pseudo initial singularity as $H\to \infty$, but pushed to $t\to \infty$. The trajectory shows that the universe performs a quasi-de sitter inflationary behaviour, then it ends at the trajectory cuts the zero acceleration curve. This exit is followed with a Type II (sudden) singularity as $\dot{H}\to \pm\infty$ at finite $H=\frac{1}{12}\sqrt{\frac{-15R_i}{\alpha}}$. In both cases, we expect that the GR to lead the evolution at relatively low Hubble regime just as in the standard cosmology.}
\label{Fig:phasespace}
\end{center}
\end{figure}
\begin{equation}\label{sol_J}
    t=t_i+24\frac{\alpha}{R_i}(H-H_i)-9\sqrt{\frac{3\alpha}{R_i}}
    \left[\arctanh{\left(2\sqrt{\frac{3\alpha}{R_i}}H\right)}-\arctanh{\left(2\sqrt{\frac{3\alpha}{R_i}}H_i\right)}\right].
\end{equation}
Although the above solution is exact, it is hard to understand the dynamical evolution of the system or to show how sensitive it is to the initial conditions. However, we can rewrite the solution in the form
\begin{equation}\label{phasespace}
\dot{H}=\frac{R_i(12\alpha H^2 -R_i)}{6\alpha(48 \alpha H^2 +5 R_i)},
\end{equation}
which represents a one dimensional autonomous system, i.e. $\dot{H}=\mathcal{F}(H)$. Consequently, we can interpret the above differential equation as a vector field on a line, whereas its trajectory in the ($H,~\dot{H}$) phase space can be used as a basic tool to analyze the dynamics of the model. In general, these systems are fully explained by the asymptotic behaviour of the trajectory $\dot{H}(H)$ and their fixed points.

It is obvious that the phase portrait \eqref{phasespace} asymptotically fixes to $\dot{H}=\frac{R_i}{24 \alpha}$, which could be within $\dot{H}<0$ or $\dot{H}>0$ regions depends on the sign of $\alpha$. Therefore, we draw the phase portrait \eqref{phasespace} in Fig. (\ref{Fig:phasespace}). For positive/negative values of $\alpha$, the model performs an inflationary behaviour at large Hubble regime with $\dot{H}<0$ (i.e Hubble decreases with time). In the following treatments, we restrict our analysis to positive values of $\alpha$. As shown in the figure, the universe has a fixed point at $H_f=\frac{1}{6}\sqrt{\frac{3R_i}{\alpha}}$ whereas $\dot{H}=0$.
\begin{figure}
\begin{center}
\includegraphics[scale=0.4]{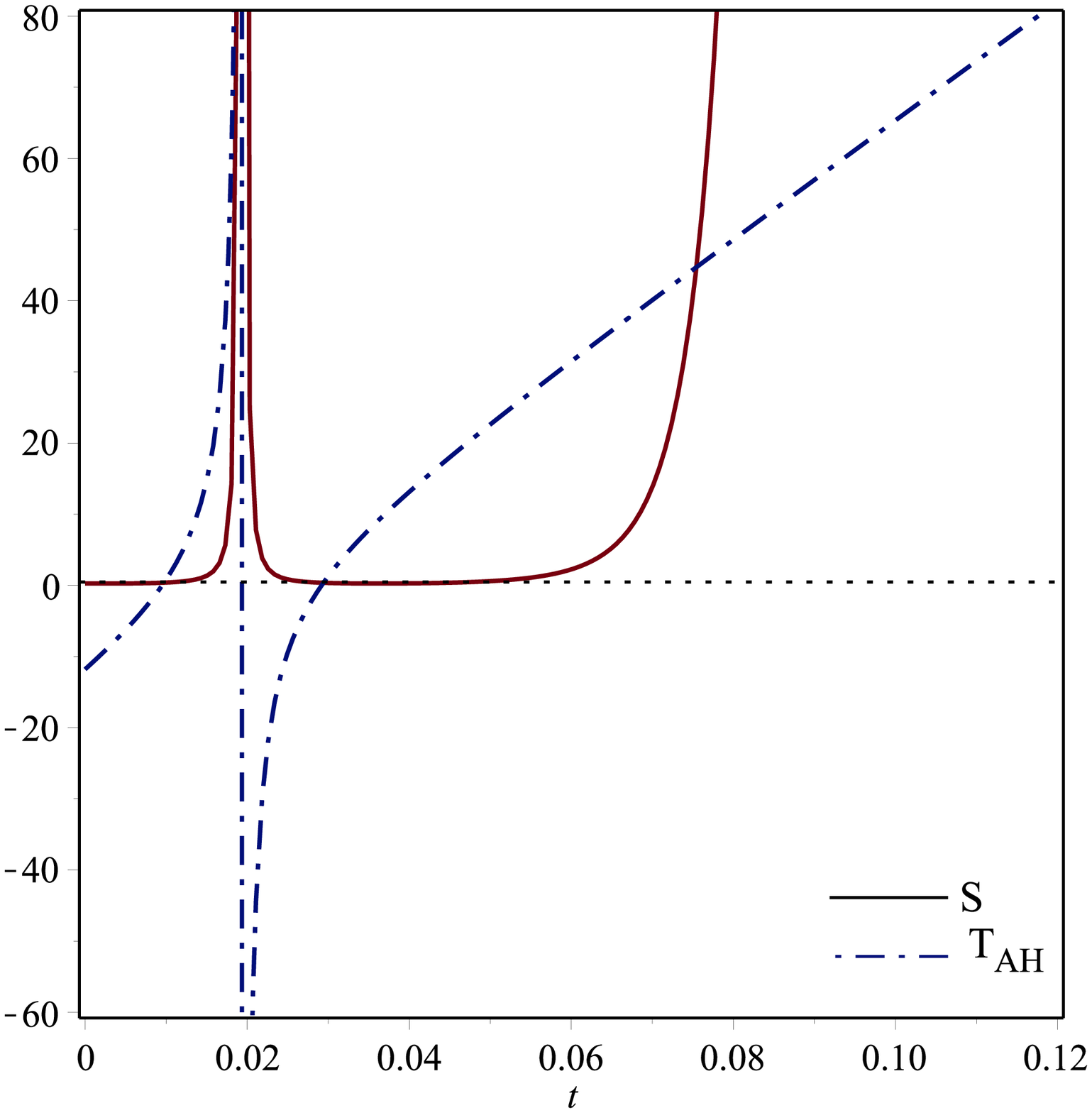}
\caption{A schematic plot of the entropy \eqref{entropy_J} and Hawking temperature \eqref{temp_J} on the apparent horizon in Jordan frame.}
\label{Fig:ST_Jor}
\end{center}
\end{figure}

The phase space analysis of the exact solution ensured the capability of the model to perform an inflationary dynamics at large Hubble regimes. In order to proceed the discussion of the thermodynamical quantities in Jordan frame, we need to obtain the Hubble parameter. From \eqref{sol_J}, we can give an approximate expressions of
\begin{eqnarray}
H &\simeq& H_i-\frac{R_i}{30\alpha}(t-t_i)\,,
\label{Hub_J} \\
a &\simeq& a_i \exp \left[ H_i (t-t_i)-(R_i/60\alpha)(t-t_i)^2
\right]\,,
\label{scf_J} \\
R &\simeq& 12H^2-R_i/(5\alpha)\,,
\label{Rsc_J}
\end{eqnarray}
where $H_i$ and $a_i$ are the Hubble parameter and the scale factor at the moment of inflation ($t=t_i)$, respectively. Then, we can directly derive the apparent horizon
\begin{equation}\label{app_hor_J}
    \mathcal{R}_{AH}=\frac{1}{H}\simeq \frac{1}{H_i-(R_i/30\alpha)(t-t_i)},
\end{equation}
and the area of the apparent horizon
\begin{equation}\label{area_hor_J}
    \mathcal{A}=4\pi \mathcal{R}^2_{AH} \simeq \frac{4\pi}{\left[H_i-\frac{R_i}{30\alpha}(t-t_i)\right]^2}.
\end{equation}
Also, we evaluate the entropy and Hawking temperature at the horizon,
\begin{eqnarray}
    S&=&\frac{1}{4} \mathcal{A} f_R \simeq \frac{12\pi\alpha\left[(t-t_i)^2 R_i^2+60\alpha (1-(t-t_i)H_i) R_i+900 \alpha^2 H_i^2\right]e^{\frac{(t-t_i)^2 R_i^2 - 15 \alpha(1+4(t-t_i)H_i)R_i+900\alpha^2 H_i^2}{75\alpha R_i}}}{R_i\left[R_i(t-t_i)-30\alpha H_i\right]^2},\label{entropy_J}\\
    T_{AH}&=&\frac{-1}{2 \pi \mathcal{R}_{AH}}\Big(1-\frac{d\mathcal{R}_{AH}/dt}{2H \mathcal{R}_{AH}}\Big)
    \simeq\frac{(t-t_i)^2 R_i^2-15\alpha\left[1+4(t-t_i)H_i\right]R_i+900\alpha^2 H_i^2}{60\pi\alpha\left[(t-t_i)R_i-30\alpha H_i\right]}.\label{temp_J}
\end{eqnarray}
In Fig. \ref{Fig:ST_Jor}, we plot Eq. \eqref{entropy_J} and \eqref{temp_J} to visualize the evolution of the entropy and Hawking temperature at the apparent horizon.

In the following subsection, we consider the same theoretical framework in Einstein frame. Also, we focus on the thermodynamical quantities which are related to the Hubble rate.
%%%%%%%%%%%%%%%%%%%%%%%%%% Section 6.2 %%%%%%%%%%%%%%%%%%%%%%%%%%%%%
\subsection{Thermodynamics in the Einstein frame}\label{S6.2}
\begin{figure}
\begin{center}
\includegraphics[scale=0.4]{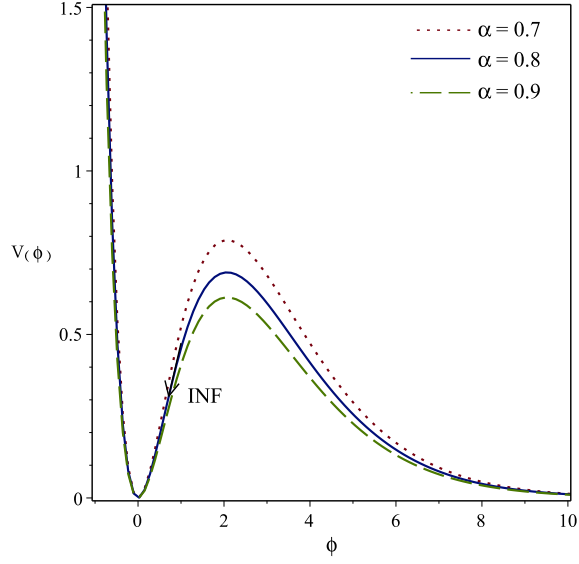}
\caption{A schematic plot of the potential \eqref{UVpot}. The potential has a global minimum at $\phi=0$, where the scalar field is normalized to Planck's mass.}
\label{Fig:potential}
\end{center}
\end{figure}

%%%%%%%%%%%%%%%%%%%%%%%%%%%%%%%%%%%%%%%%%%%%%%%%%%%%%%%%%%%%%%%%%%%%
Let us consider inflationary dynamics in the Einstein frame for the model~(\ref{UV_fRinf}) in the absence of matter fluids. The field $\phi$ in the Einstein frame corresponds to (\ref{UV_fRinf}) is defined by
\begin{equation}\label{UVscalarf}
\phi=\sqrt{\frac{3}{2\kappa}} \ln f_R= \sqrt{\frac{3}{2\kappa}}\,
\ln \left[ \alpha\frac{R}{R_i}+\ln\left(1+\alpha\frac{R}{R_i}\right) \right]\,.
\end{equation}
Using this relation, the field potential \eqref{pot-fR} reads
\begin{equation}\label{UVpot}
V(\phi)=\frac{R_i}{2\alpha\kappa} e^{-\sqrt{\frac{2\kappa}{3}}\phi}\left[1-\mathcal{W}\left(e^{1+\sqrt{\frac{2\kappa}{3}}\phi}\right)\right]^2
\frac{1}{\mathcal{W}\left(e^{1+\sqrt{\frac{2\kappa}{3}}\phi}\right)}\,,
\end{equation}
where $\mathcal{W}(x)$ denotes the Lambert-W function, which is the solution of the transcendental equation $W e^W = x$. The potential drops to zero at $\phi=0$, which allows the scalar field to oscillate about its global minimum and perform a reheating process. We plot the potential pattern in Fig. \ref{Fig:potential}. In general this pattern is compatible with inflationary potential. We note that the slow-roll analysis and the reheating process should be applied in details\footnote{This work is in progress.}. However, we focus in the present work on the thermodynamical quantities in Einstein frame.
%[\textbf{The potential pattern needs to briefly discussed.}]%For discussion see \cite{Martin:2013tda}, Sec 6.5 Logarithmic potential inflation (LPI).

The relation between the cosmic time $\bar{t}$ in the Einstein frame and that in the Jordan frame is given by $d\bar{t}=\Omega\, dt$. Recalling Eqs.~\eqref{Hub_J} -- \eqref{Rsc_J}, we write
\begin{equation}
\bar{t} = \int_{t_i}^t \Omega\, dt \simeq \frac{2}{5}e^{\frac{60\alpha H_i^2-R_i}{10 R_i}} \left[ C_1 (t-t_i) +C_2 (t-t_i)^2 \right]\,,\label{tiltre}
\end{equation}
where
$$C_1=2/5\,{{ e}^{1/10\,{\frac {-{ R_i}+60\,{{ H_i}}^{2}\alpha}{{
R_i}}}}}\sqrt {5}\sqrt {15\,{{ H_i}}^{2}\alpha+{ R_i}}{\frac {1}{
\sqrt {{ R_i}}}},$$
$$C_2=-6/5\,{{ e}^{1/10\,{\frac {-{ R_i}+60\,{{ H_i}}^{2}\alpha}{{
 R_i}}}}}\sqrt {5} \left( {\frac {3}{20}}\,{ R_i}+{{ H_i}}^{2}
\alpha \right) { H_i}{\frac {1}{\sqrt {15\,{{ H_i}}^{2}\alpha+{
 R_i}}}}{\frac {1}{\sqrt {{ R_i}}}}.$$
where $t=t_i$ corresponds to $\bar{t}=0$. Also, the Einstein frame scale factor, $\bar{a}(\bar{t})=\Omega\, a(t)$, takes the form
\begin{equation}
\bar{a}(\bar{t}) \simeq \bar{a_i} \left[\frac{5}{2}\alpha H_i^2+\frac{1}{6}R_i - \left(\alpha H_i^2+\frac{3}{20}R_i\right)H_i\bar{t}\right] \,\exp\left(H_i \bar{t}-\frac{R_i}{60\alpha}\bar{t}^2+\frac{6\alpha H_i}{R_i}-\frac{1}{10}\right)\,,
\label{aein}
\end{equation}
where $\bar{a}_i= \frac{12 a_i}{\sqrt{5R_i(15\alpha H_i^2+R_i)}}$. Similarly the evolution of the Hubble parameter,
$\bar{H}=\frac{1}{\bar{a}}\left(d\bar{a}/d\bar{t}\right)=\frac{1}{\sqrt{f_R}}\left(H+\frac{\dot{f_R}}{2 f_R}\right)$,
is given by
\begin{equation}
\bar{H}(\bar{t}) \simeq \frac{30\alpha(90\alpha H_i^2+R_i)H_i-10(R_i^2+42\alpha R_i H_i^2 +180 \alpha^2 H_i^4)\bar{t}+3(20\alpha H_i^2+3R_i)R_i H_i\bar{t}^2}{30\alpha\left[10(15\alpha H_i^2+R_i)-3\left(20\alpha H_i^2+3R_i\right)H_i\bar{t}\right]}\,.
\label{Hein}
\end{equation}
\begin{figure}
\begin{center}
\includegraphics[scale=0.4]{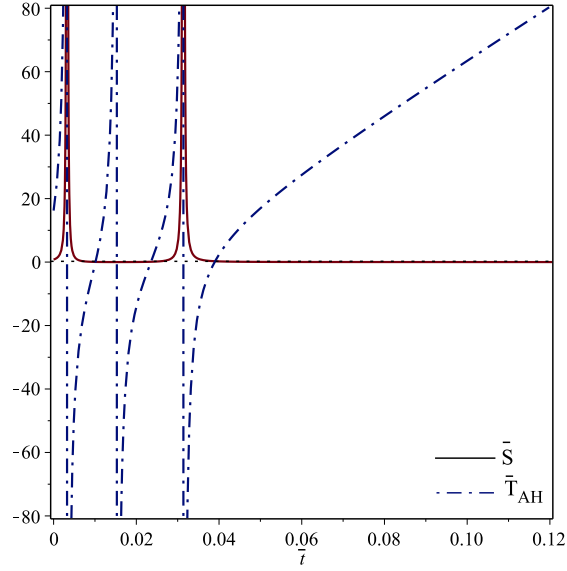}
\caption{A schematic plot of the entropy \eqref{entropy_E}, and Hawking temperature \eqref{temp_E} on the apparent horizon in Einstein frame.}
\label{Fig:ST_Ein}
\end{center}
\end{figure}

In order to evaluate the thermodynamical quantities related to the Hubble rate in Einstein frame, we obtain the apparent horizon
\begin{equation}\label{eq:app_hor1}
    \bar{\mathcal{R}}_{AH}\simeq \frac{30\alpha\left[10(15\alpha H_i^2+R_i)-3\left(20\alpha H_i^2+3R_i\right)H_i\bar{t}\right]}{30\alpha(90\alpha H_i^2+R_i)H_i-10(R_i^2+42\alpha R_i H_i^2 +180 \alpha^2 H_i^4)\bar{t}+3(20\alpha H_i^2+3R_i)R_i H_i\bar{t}^2}.
\end{equation}
Then, the area of the cosmological horizon is
\begin{equation}\label{eq:area_hor1}
    \bar{\mathcal{A}}=4\pi \bar{\mathcal{R}}_{AH}\simeq \frac{120\alpha\pi\left[10(15\alpha H_i^2+R_i)-3\left(20\alpha H_i^2+3R_i\right)H_i\bar{t}\right]}{30\alpha(90\alpha H_i^2+R_i)H_i-10(R_i^2+42\alpha R_i H_i^2 +180 \alpha^2 H_i^4)\bar{t}+3(20\alpha H_i^2+3R_i)R_i H_i\bar{t}^2}.
\end{equation}
Now, we write the apparent horizon entropy and Hawking temperature in Einstein frame, respectively, as below
\begin{equation}\label{entropy_E}
    \bar{S}=\frac{1}{4}\bar{\mathcal{A}}\simeq \frac{30\alpha\pi\left[10(15\alpha H_i^2+R_i)-3\left(20\alpha H_i^2+3R_i\right)H_i\bar{t}\right]}{30\alpha(90\alpha H_i^2+R_i)H_i-10(R_i^2+42\alpha R_i H_i^2 +180 \alpha^2 H_i^4)\bar{t}+3(20\alpha H_i^2+3R_i)R_i H_i\bar{t}^2},
\end{equation}
\begin{eqnarray}\label{temp_E}
    \bar{T}_{AH}&=&\frac{-1}{2 \pi \bar{\mathcal{R}}_{AH}}\Big(1-\frac{d\bar{\mathcal{R}}_{AH}/d\bar{t}}{2\bar{H} \bar{\mathcal{R}}_{AH}}\Big) \simeq \frac{1}{216000\alpha\pi}
\nonumber \left[\left( 1080\,\alpha\,{R_{i}}^{3}H_{i}^{4}+3600\,{\alpha}^{2}{R_{i}}^{2}H_{i}^{6}+81\,{R_{i}}^{4}H_{i}^{2} \right) \bar{t}^{4}- \left( 180\,{R_{i}}^{4}H_{i}\right.\right.\\
\nonumber &&\left.+8760\,\alpha\,{R_{i}}^{3}H_{i}^{3}+82800\,{\alpha}^{2}{R_{i}}^{2}H_{i}^{5}+ 216000\,{\alpha}^{3}R_{i}\,H_{i}^{7}\right) \bar{t}^{3}+ \left(7725\,{R_{i}}^{3}H_{i}^{2}\alpha+248400\,H_{i}^{4}{\alpha}^{2}{R_{i}}^{2}+100\,{R_{i}}^{4}\right.\\
\nonumber&&\left.+1782000\,H_{i}^{6}{\alpha}^{3}R_{i}+3240000\,H_{i}^{8}{\alpha}^{4} \right) \bar{t}^{2}+ \left(-2106000\,{\alpha}^{3}R_{i}\,H_{i}^{5}-9720000\,{\alpha}^{4}H_{i}^{7}-20700\,{\alpha}^{2}{R_{i}}^{2}H_{i}^{3}\right.\\
\nonumber&&\left.\left.+2100\,R_{i}^{3}H_{i}\,\alpha \right)\bar{t}
-80550\,H_{i}^{2}{\alpha}^{2}{R_{i}}^{2}-1500\,\alpha\,{R_{i}}^{3}+5670000\,H_{i}^{6}{\alpha}^{4}-661500\,H_{i}^{4}{\alpha}^{3}R_{i}
\right]/\\
\nonumber&&\left[\left(\left( {\frac {3}{20}}\,H_{i}\,{R_{i}}^{2}+{H_{i}}^{3}\alpha
\,R_{i} \right) \bar{t}^{2}+ \left( -\frac{1}{6}\,{R_{i}}^{2}-7\,\alpha\,R_{i}\,{H_{i}}^{2}-30\,{H_{i}}^{4}{\alpha}^{2} \right) \bar{t}
+\frac{1}{2}\,H_{i}\,R_{i}\,\alpha+45\,{H_{i}}^{3}{\alpha}^{2}
\right)\times\right. \\
&&\left.\left(\left( {\frac {3}{20}}\,R_{i}\,H_{i}+{H_{i}}^{3}\alpha
 \right) \bar{t}-\frac{5}{2}\,{H_{i}}^{2}\alpha-\frac{1}{6}\,R_{i}
\right)\right].
\end{eqnarray}
In Fig. \ref{Fig:ST_Ein}, we plot Eqs. \eqref{entropy_E} and \eqref{temp_E} to visualize the evolution of the entropy and Hawking temperature at the apparent horizon in Einstein frame. The thermodynamical quantities in Einstein frame are clearly different from those in Jordan frame, which shows that the two frames are not equivalent on the thermodynamical level.

%%%%%%%%%%%%%%%%%%%%%%%%%%%%%%%%%%%%%%%%%%%%%%%%%%%%%%%%%%%%%%%%%%%%%%%%%%%%%%%
\section{Conclusions}\label{S7}

In the line of research on the (in)equivalence of Jordan and Einstein frames, we investigated this issue within the $f(R)$ gravity framework on the thermodynamic level. Therefore, for several $f(R)$ gravity theories, we calculated some thermodynamic quantities: Bekenstein-Hawking entropy, Hawking temperature, the heat capacity, the quasi-local energy and Gibbs energy.

First, we derived some vacuum solutions, in the presence of electromagnetism, of a spherically symmetric black hole: For the $f(R)=R+\alpha R^2$ gravity, we derived the solution in Jordan frame which gives a constant Ricci scalar, $R=-8\Lambda$. In this case, the conformal factor $\Omega=\sqrt{1-16\alpha\Lambda}$ is constant and consequently all the thermodynamic quantities in Einstein frame are the same as in Jordan frame. For the newly introduced $f(R)=R+2\alpha \sqrt{R+8\Lambda}$ gravity, we derived the solutions in Jordan frame which gives a non-constant Ricci scalar, $R=-8\Lambda+\frac{1}{r^2}$. In this case, the conformal factor has a radial dependance $\Omega=\sqrt{1+\alpha r}$ and consequently all the thermodynamic quantities are nontrivially modified. Therefore, as we demonstrated, for the black hole solutions with constant Ricci scalars, the thermodynamical quantities in the two frames are equivalent. However, for the case of black holes with non-constant Ricci scalars, the thermodynamical equivalence of the two frames is no longer valid.

Second, we derived some cosmological solutions with an FLRW background: For Fonarev solution, which is characterized by the exponential scalar field potential, in Einstein frame, we evaluated some thermodynamic quantities that are related to Hubble rate. In order to examine the thermodynamic in Jordan frame, we derived   the corresponding $f(R)\propto R^n$ which in return produces a power-law cosmology. In this case, we found that the thermodynamic in the two frames is equivalent. For the newly introduced $f(R)=R e^{\alpha R/R_i}$ UV gravity, we derived the Jordan frame solution and then some thermodynamic quantities related to Hubble rate. Then, we derived the corresponding scalar-tensor theory in Einstein frame, which introduces a new inflationary potential. As we showed, the thermodynamic quantities in Einstein frame are nontrivially modified. We believe that the accidental equivalence of the two frames in the context of the power-law cosmology, may be explained by studying in depth the correspondence of the dynamical systems corresponding to the two frames, and thus relating the trajectories in the phase space, or even perhaps the fixed points.

In conclusion, $f(R)$ gravity and its corresponding scalar-tensor theory are mathematically equivalent, at least when conformal invariant quantities are considered. However, this equivalence is clearly broken when singularities are present as it has been explicitly demonstrated in Ref. \cite{Briscese:2006xu}. As we demonstrated, the two frames are not thermodynamically equivalent at a quantitative level, in terms of several physical quantities. This clearly indicates that the notion of frame-independent quantities as real physical quantities should be further developed.

\begin{acknowledgments}

This work is supported by MINECO (Spain), FIS2016-76363-P, and by
project 2017 SGR247 (AGAUR, Catalonia) (S.D.O). This work is
supported by the DAAD program ``Hochschulpartnerschaften mit
Griechenland 2016'' (Projekt 57340132) (V.K.O). V.K.O is indebted
to Prof. K. Kokkotas for his hospitality in the IAAT, University
of T\"{u}bingen.
\end{acknowledgments}

%%%%%%%%%%%%%%%%%%%%%%%%%%%%%%%%%%%%%%%%%%%%%%%%%%%%%%%%%%%%%%%%%%%%%%%%%%%%%%%%%%%%%%
%\bibliographystyle{apsrev}
%\bibliography{DK12407R1}

\end{document}